\documentclass[prd, aps,nofootinbib,preprintnumbers]{revtex4}
\usepackage{amsmath, amssymb, graphicx, epsf}
\usepackage{epstopdf}
\usepackage{graphicx}
\usepackage{dcolumn}
\usepackage{bm}
\usepackage{hyperref}
\newcommand{\md}{\mathrm{d}}
\newcommand{\mpl}{M_{\mathrm{p}}}
\newcommand{\nc}{\newcommand}
\nc{\ba}{\begin{eqnarray}}
\nc{\ea}{\end{eqnarray}}
\newcommand\be{\begin{equation}}
\nc{\K}{{\bf k }}
\nc{\cn}{\mathrm{cn(x|\frac{1}{2})}}

\begin{document}
\title{Tachyonic Resonance Preheating in Expanding Universe
}
\author{Ali Akbar Abolhasani$^{1,2}$}
\email{abolhasani(AT)ipm.ir}
\author{Hassan Firouzjahi$^{2}$}
\email{firouz(AT)ipm.ir}
\author{M. M. Sheikh-Jabbari$^{2}$}
\email{jabbari(AT)theory.ipm.ac.ir}
\affiliation{$^{1}$
Department of Physics, Sharif University of Technolog, Tehran, Iran}
\affiliation{$^{2}$ School of Physics, Institute for Research in Fundamental Sciences (IPM),
P. O. Box 19395-5531,
Tehran, Iran}

\date{\today}

\begin{abstract}

In this paper the tachyonic resonance preheating generated from the bosonic trilinear $\phi\chi^2$ interactions in an expanding Universe is studied. In $\lambda\phi^4/4$ inflationary model the trilinear interaction, in contrast to the four-legs $\phi^2\chi^2$,
breaks the conformal symmetry explicitly and the resonant source term becomes non-periodic, making the  Floquet theorem inapplicable.  We find that the occupation number of the produced $\chi$-particles has a non-linear exponential growth with exponent $\sim x^{3/2}$, where $x$ is the conformal time.
This should be  contrasted with preheating from a periodic resonant source, arising for example from the four-legs $\phi^2\chi^2$ interaction, where the occupation number has a linear exponential growth. We present an analytic method to compute the interference term coming from phases accumulated in non-tachyonic scattering regions and show that the effects of
the interference term causes ripples on $x^{3/2}$ curve, a result which is
confirmed by numerical analysis. Studying the effects of  back-reaction of the $\chi$-particles, we show that tachyonic resonance preheating in our model can last long enough to transfer most of the energy from the background inflation field $\phi$, providing an efficient model for preheating in the  chaotic inflation models.

\vskip 2cm\begin{center}
\textit{This work is dedicated to the  memory of Lev Kofman.}
\end{center}
\end{abstract}

\preprint{IPM/P-2009/052}

\maketitle
\section{Introduction}
At the end of inflation the Universe is extremely cold and all of its energy is concentrated in the inflaton field $\phi$. Therefore, the reheating stage is a key ingredient to connect a successful
inflationary stage into hot big bang cosmology. In simplest models of inflation, such as chaotic inflation models, inflaton field oscillates coherently around the minimum of the potential with amplitude close to the Planck mass, $M_P$.\footnote{We use the convention in which $M^2_P=1/G_N$, for $G_N$ being the Newton constant.} To transfer the energy from the background inflaton field and reheat the Universe  one has to consider its coupling to other fields, such as the  Standard Model fields. Due to coherent oscillations of the background inflaton field around
the potential minimum, explosive particle creation can happen via parametric resonance which can drag most of the energy from the inflaton field and reheat the Universe \cite{Kofman:1997yn, Kofman:1994rk, Traschen:1990sw, Shtanov:1994ce}. Preheating, the stage of explosive particle creation via parametric resonance, is a non-perturbative effect and
the produced particles are highly non-thermal. Subsequent turbulent interactions of different modes and re-scatterings
would bring an end to preheating and thermalization starts \cite{Khlebnikov:1996mc, Prokopec:1996rr,  Berges:2002cz,
Felder:2000hr, Podolsky:2005bw, Micha:2004bv} where the perturbative reheating mechanism may start
\cite{Dolgov:1982th, Abbott:1982hn}.  For a review of preheating see \cite{Bassett:2005xm} and references therein.

There have been many studies of preheating via four-legs interaction $g^2 \phi^2 \chi^2$
between the inflaton field and the resonance field $\chi$
in  expanding or  flat backgrounds.  One disadvantage of  $g^2 \phi^2 \chi^2$ preheating channel is that the decay of inflaton in an expanding background is not complete and the inhomogeneous  inflaton particles will dominate with the matter equation of state, an unsatisfactory state to end (p)reheating with.

In order to enhance the preheating mechanism other interactions are
necessary, specially once the inflaton field amplitude is damped while
oscillating around its minimum. The simplest and the most natural of these
interactions is the trilinear interaction $\sigma \phi \chi^2$ where
$\sigma$ is a constant of dimension of mass. The implications of
trilinear interaction for the chaotic inflationary potential
$m^2 \phi^2/2$ was studied in \cite{Dufaux:2006ee}. One interesting aspect
of preheating via trilinear interaction is that during half period of
its oscillation, the inflaton field changes sign and
the time-dependent frequency squared for the $\chi$ particle creation
becomes negative for small enough mode wavelengths.
This tachyonic  instability enhances  the preheating mechanism
significantly  and  in \cite{Dufaux:2006ee} was dubbed as
``tachyonic resonance''. The analysis of \cite{Dufaux:2006ee} were mainly
devoted to a flat background and  the effects of tachyonic resonance in
an expanding background was briefly considered, incorporating them
``adiabatically''. The tachyonic effect of a trilinear interaction
was used in \cite{Shuhmaher:2005mf},  in the context of cosmological moduli problem and the idea of tachyonic parametric resonance with a negative coupling constant was also exploited in
\cite{Greene:1997ge}.

Here we consider the tachyonic resonance preheating in $\lambda \phi^4/4$ and $m^2 \phi^2/2$  inflationary potentials in an expanding Universe. Our main interest will be  in the $\lambda \phi^4/4$ theory but we extend the results of \cite{Dufaux:2006ee} for $m^2 \phi^2/2$ inflationary model to an expanding background.

The preheating with four-leg interaction within the $\lambda \phi^4/4$ inflation theory was studied extensively in \cite{Greene:1997fu}. Due to the conformal invariance of the inflationary potential, the effects of expanding background can be absorbed by a conformal transformation and one is basically dealing with preheating in a Minkowski background. However, with the addition of $\sigma \phi \chi^2$  interaction, the conformal invariance is broken explicitly and the methods  employed in \cite{Greene:1997fu} should now be modified. Breaking of conformal invariance results in an explicit breaking of periodicity in $\chi$-modes equation: the amplitude of the source term in $\chi$ field particle creation grows linearly with the dimensionless conformal time $x$. Consequently, the standard Floquet theorem does not apply. This leads to the interesting result that the number density of  $\chi$ particles produced at momentum $\K$, $n_\K$, grows as $e^{c\,  x^{3/2}}$ for some constant $c$. This should be compared with the standard result of the preheating due to a periodic source term, where $n_\K$ has a linear exponential growth as dictated by the Floquet theorem.

Another interesting result of our analysis is that we present an
analytic method to compute the interference term which accounts for
the phase difference between the $\chi$ fields during each period of
the $\phi$ field oscillation. The phase accumulated during certain
number of oscillations, which we compute, makes the $\chi$ particle
creation to be destructive for certain modes, the $k^2$ of which we specify.
This causes some ripples
on the $x^{3/2}$ curve in the $\ln n_\K$ diagram.
This result is to be compared with the stochastic nature of the phases in the four-leg preheating scenarios \cite{Kofman:1997yn,Greene:1997fu}.

While our main interest is in $\lambda \phi^4/4$ inflationary theory,
we also study the effect of expanding background on $m^2 \phi^2/2$ theory.
As one may expect, the effect of expanding background will generally
suppress the preheating efficiency. As we demonstrate, however,  it
actually enhances the preheating for some certain modes.

As the number of $\chi$ particles grow their back-reaction on the dynamics
of the particle production and the $\phi$ field dynamics become important,
slowing down the preheating and eventually terminating it. For a successful preheating model it is necessary that this back reaction does not become large too early, before the energy of the inflaton field
is completely transferred into the $\phi$ or $\chi$ particles. It is also
important that in the end of preheating we remain with a relativistic ensemble of these particle. As in the case of $m^2\phi^2$, which has been analyzed in \cite{Dufaux:2006ee}, our analysis shows that the trilinear interactions seems to lead to a more efficient preheating than the four-leg case.

The paper is organized as follows. In section \ref{phi4-case} we study in details the tachyonic resonance for $\lambda \phi^4/4$ theory. We demonstrate that our analytical results agree very well with the exact numerical results.  In section \ref{m2phi2-case} we repeat the analysis of tachyonic resonance for $m^2 \phi^2/2$ theory in an expanding background.
In section \ref{back-reaction} we study the effects of $\chi$-particles  back-reactions
on the preheating and estimate the time preheating ends.
A summary of the results are given in section \ref{summary}.
Some technical aspects of the analysis are relegated into
{\bf Appendices \ref{acn}-\ref{r+jacobicn}}.

\section{The Tachyonic Resonance in $\lambda \phi^4/4$ theory}
\label{phi4-case}

As explained above, we are mainly interested in tachyonic resonance in $\lambda \phi^4/4$ inflationary
theory with the trilinear interaction $\sigma \phi \chi^2$ where $\sigma$ is a parameter
with dimension of mass. As in \cite{Dufaux:2006ee} we have to include the self-interaction $\lambda' \chi^4$ to uplift the potential and keep it bounded from below. The total potential is
\ba
\label{V-eq}
V&=& \frac{\lambda}{4} \phi^4 + \frac{\sigma}{2} \phi \chi^2 + \frac{\lambda'}{4} \chi^4 +
\frac{\sigma^4}{16 \lambda \lambda'^2} \\
&=& \lambda \left( \frac{\phi^2}{2} - \frac{\sigma^2}{4\lambda \lambda'}  \right)^2
+ \frac{\lambda'}{4} \left( \chi^2+ \frac{\sigma \phi}{ \lambda'}  \right)^2 \, .
\ea
The potential has a global minimum at $\phi_0= -\sqrt{ \sigma^2/2\lambda \lambda'}$
and $\chi_0^2 = - \sigma \phi_0/\lambda'$. The last term in (\ref{V-eq}) is added to lift
the global minimum to zero. In order to make sure that this constant value does not contribute to the inflationary dynamics, we require that $\sigma^4/ \lambda \lambda'^2 \ll \lambda \phi_{end}^4$ where $\phi_{end}$ is the value of the inflaton field at the end of inflation when the onset of preheating starts with $\phi_{end} \simeq M_P/\sqrt{\pi}$. This leads to
\ba
\label{cond1}
\frac{\sigma}{ \sqrt{\lambda \lambda'} M_P } \ll 1 \, .
\ea

If we assume that the $\chi$-field is heavy during inflation and is settled down to its minimum at $\chi_0$ this requires that $12 \sigma M_P^2/\lambda \phi^3 >1$, which can be satisfied if
$ 12 \sigma M_P^2/\lambda \phi_i^3 >1$ where $\phi_i$ is the initial value of the inflaton field at the start of inflation. For inflation to solve the flatness and the horizon problem, we require that $\phi_i \simeq 10 M_P$ and $\lambda \simeq 10^{-14}$. So the assumption that
the field $\chi$ would be stabilized to its minima requires that
\ba
\label{cond2}
 \frac{\sigma}{ \lambda  M_P }
\gtrsim 10^2 \, .
\ea
However, it is also possible that  the field $\chi$ is light during inflation
which can then contribute to iso-curvature perturbations.

To study the preheating with trilinear interaction, we review the background of \cite{Greene:1997fu} where the parametric resonance with four-legs interaction for $\lambda \phi^4/4$ theory was investigated. As in \cite{Greene:1997fu}, performing the conformal transformation
$\varphi \equiv a (\eta) \phi$ the background equation for the $\varphi$ field to a very good approximation is simplified to
\ba
\label{phi-eq}
\varphi'' + \lambda \varphi^3 =0 \, ,
\ea
where the derivatives are with respect to the conformal time $d\eta = dt/ a(t)$  where $a(t)$ is the scale factor. The solution to this equation is given in terms of Jacobi elliptic cosine function
\ba
\varphi = \tilde \varphi f(x) \, ,
\ea
where $\tilde \varphi$ is the amplitude of the oscillations at the onset of preheating
which we take to be $0.1 M_P$, and
the dimensionless conformal time $x$ is defined by
\ba
\label{x-def}
x \equiv \sqrt{\lambda} \tilde \varphi \eta = \left(   \frac{6 \lambda M_P^2}{\pi}  \right)^{1/4}
\sqrt{t} \, .
\ea
The Jacobi elliptic cosine function satisfies the relation $f'^2 = \frac{1}{2} (1 - f^4)$ with the solution denoted by \footnote{Our definition of Jacobi elliptic cosine conforms to that of
\cite{abramowitz} which is slightly different than that of \cite{Greene:1997fu}. }
\ba
f(x) = cn(x, \frac{1}{2}) \, .
\ea
The function
$f(x)$ is a periodic function with the periodicity
\ba\label{periodT}%
T\equiv 4 K(1/2) \simeq 7.416
\ea%
where $K(m)$ represents the complete elliptic integral of the first kind. For a review of Jacobi elliptic cosine function see {\bf Appendix \ref{acn}}.

One also notes that the following relations also hold
\ba
\label{background}
a(\eta) = \sqrt{\dfrac{2\pi\lambda}{3}}\dfrac{\tilde{\varphi}^2}{M_p}\eta= \sqrt{\dfrac{2\pi}{3}}\dfrac{\tilde{\varphi}}{M_p}\ x \ ,  \qquad  t = \sqrt{\dfrac{\pi\lambda}{6}}\dfrac{\tilde{\varphi}^2}{M_p}\eta^2 \, .
\ea

Ignoring the back-reaction of the produced $\chi$ particles, performing the conformal transformation $  \hat \chi = a(\eta) \chi$, and using the background equations (\ref{background}), the equation for the mode function $\chi_\K$ in
the momentum space is
\ba
\label{X-eq}
\hat\chi''_\K + \left( \kappa^2 + p\, x\, f(x) \right) \hat \chi_\K =0 \, ,
\ea
where the derivatives are now with respect to the dimensionless conformal time $x$ given
in (\ref{x-def}), and
\ba
\kappa^2=\frac{\K^2}{\lambda\tilde\varphi^2} ,\qquad p \equiv \sqrt{\dfrac{2\pi}{3}}\left(\dfrac{\sigma}{\lambda M_p}\right) \, .
\ea
In writing \eqref{X-eq} we have ignored the $\chi_0$ and $\phi_0$ terms, associated with the minima of the potential
\eqref{V-eq}. The effects of non-zero values for $\chi_0$ and $\phi_0$  will appear in \eqref{X-eq} as the shift in
$p$  by an amount of the order $p\frac{\phi_0}{\tilde\varphi}$ (and similarly for the $\kappa^2$ term), causing
 an error of the order $p\sqrt{\frac{\lambda}{\lambda'}}$, which  using \eqref{cond1} it is much smaller than one.

Before we start the analytical theory of tachyonic resonance, we need to estimate the magnitude of parameter $p$. This in turn is determined by the conditions whether
the field $\chi$ is heavy or light during inflation.  If we assume that the field $\chi$ is heavy during inflation, then the condition (\ref{cond2}) indicates that $p \ge 100$. However, if
we assume that the field $\chi$ is light during inflation, which is our preferred choice, then (\ref{cond1}) can be used
to estimate the bound on $p$. Writing (\ref{cond1}) in the form
$p \sqrt{\lambda/\lambda'} \ll 1$, we see that for natural choice of $\lambda \sim \lambda'$,
$p \ll 1$. On the other hand, for $\lambda' \gg \lambda$, the bound on
$p$ may become relaxed.  We present our analytical results for arbitrary value of $p$. However, for numerical examples we shall consider the cases  $p\ll 1$ and $p \sim 1$ for illustrations.

\begin{figure}[t]
\centerline{\includegraphics[width=10cm]{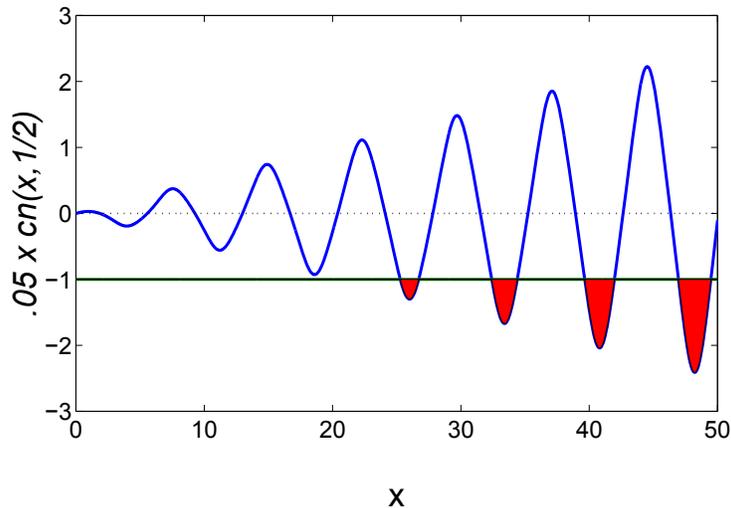}}
\caption{\label{rfx}
A plot of $p\, x\, f(x)$ given by  (\ref{omega}) is presented here for $p=0.05$. The horizontal line represents the solutions of $\omega^2 (x)= \kappa^2 + p \, x f(x) =0$ for $\kappa^2=1$.
The filled regions shows the tachyonic regions where $\omega^2(x) <0$. In our notation, the $ j^{th} $  tachyonic region is confined to $x^-_j < x < x^+_j$.}
\end{figure}

Equation (\ref{X-eq}) represents a  harmonic oscillator with the time dependent frequency
\ba
\label{omega}
\omega^2 (x) \equiv  \kappa^2 + p \, x f(x) = \kappa^2 + p \, x \, cn(x, \frac{1}{2}) \, .
\ea
A plot of $p \, x\,  f(x)$ is given in {\bf Fig. \ref{rfx}}. We see that $\omega^2(x)$ is not periodic and its maximum value is increasing linearly with the conformal time $x$. Consequently, the Floquet theorem for
particle creation via parametric resonance with a periodic source does not apply. We need to employ the direct scattering method to evaluate the number density of the
$\chi$ particle, $n_\K$. For a given $\kappa^2\neq 0$,  $\omega^2(x)$ is initially positive. However, for $t>t_\star$ it becomes tachyonic and the method of tachyonic resonance considered in \cite{Dufaux:2006ee} would apply. Interestingly, the zero momentum case, $\kappa=0$, is tachyonic for all the time. We therefore start our analytical study of tachyonic resonance
with this simple case and defer the case of $\kappa^2 \neq 0$ to the section
\ref{non-zero-k}.

\subsection{Tachyonic Resonance with  $\kappa^2=0$}
\label{zero-k}
As can be seen from {\bf Fig. \ref{rfx}}, $\omega^2(x)$ becomes negative in the region where
$f(x)<0$. The Jacobi elliptic cosine is periodic with period $T= 4 K(1/2)$ and has roots at
\begin{equation}
\label{turningpoint}
x_{j}^{-} = (j - \dfrac{3}{4}) T \quad ,  \quad x_{j}^{+} = (j - \dfrac{1}{4}) T \, ,
\end{equation}
where the integer number $j=1,2,...$ counts the number of oscillations of $f(x)$ starting from $x=0$.  In this convention, the $ j^{th} $  tachyonic region is confined to $x^-_j < x < x^+_j$.
Similarly, the minima and the maxima of $f(x) $ are given by
\begin{equation}
x_{j}^{max} = ( j - 1) T \quad , \quad
x_{j}^{min} = (j - \dfrac{1}{2})T .
\end{equation}
In the region $x_{j}^- < x < x_j^+$, $\omega^2(x) <0$ and the method of tachyonic resonance developed in  \cite{Dufaux:2006ee} applies.

As in \cite{Dufaux:2006ee}, in the region  $\omega^2(x)>0$ and when the adiabadicity condition $|  \omega|'/\omega^2 \ll 1$ holds, the solution of  (\ref{X-eq}) can be
given in WKB approximation as
\ba
\label{WKB1}
\hat \chi_{\K=0} (x) \simeq  \frac{\alpha^j}{\sqrt{2 \omega (x)}} \exp \left( - i \int_{x_0}^x \omega (x')  d x' \right) +
\frac{\beta^j}{\sqrt{2 \omega (x)}} \exp \left( + i \int_{x_0}^x \omega (x')  d x' \right) \, ,
\ea
where $\alpha^j$ and $\beta^j$ are the Bogoliubov coefficients  with normalization
$| \alpha|^2 - |\beta|^2 =1$. At the beginning of preheating when the inflaton field starts
its oscillation toward the potential minimum, there is no $\chi$ particle and we start
with the vacuum initial condition $\alpha^0 =1$ and $ \beta^0 =0$. The occupation number of the $\chi$ particle after $j$ oscillation (or after $j$ tachyonic regions) is given by
\ba
n_{\K=0}^j = | \beta^j |^2 \, .
\ea
For the tachyonic region $x^-_j < x < x^+_j$, the WKB approximation holds again for the frequency $\Omega^2(x) \equiv -\omega^2(x)$ and the
solution is given as a superposition of exponentially growing and
decaying parts:
\ba
\label{WKB2}
\hat \chi_{\K=0} (x) \simeq  \frac{a^j}{\sqrt{2 \Omega (x)}} \exp \left( -  \int_{x_-^j}^x \Omega (x')  d x' \right) + \frac{b^j}{\sqrt{2 \Omega (x)}}
\exp \left( +  \int_{x_-^j}^x \Omega (x')  d x' \right) \, ,
\ea
where $a^j$ and $b^j$ are constants of integration. Finally, after $x> x_+^j$, the solution
is given by (\ref{WKB1}) with $j \rightarrow j+1$.

Around the  points $x_-^j$ and $x_+^j$ where the adiabatic approximation is broken, we have to solve the mode equation (\ref{X-eq}) as in
scattering theory and match it with the solutions (\ref{WKB1}) and (\ref{WKB2}). Following
the methods of \cite{Dufaux:2006ee} in performing this matching condition, we obtain
the following transfer matrix, relating the Bogoliubov coefficients after $j^{th}+1$ scattering
to those of $j^{th}$ scattering via
\begin{equation}
\begin{pmatrix}
\alpha^{j + 1} \\
\beta^{j +1}
\end{pmatrix}
= \mathrm{e}^{X^j}
\begin{pmatrix}
1 & i \mathrm{e}^{2i\theta^j} \\
-i \mathrm{e}^{-2i\theta^j} & 1
\end{pmatrix}
\begin{pmatrix}
\alpha^{j } \\
\beta^{j }
\end{pmatrix}
\end{equation}
where
\ba
\label{Xj}
X^j = \int _ {x_-^j} ^{x_+^j} \Omega(z) \, \mathrm{d}z \, ,
\ea
and $\theta^j $ is the total phase accumulated from $x_0$ to $x^-_j$ during the non-tachyonic intervals where $\omega^2 > 0 $. In our case, we have $\theta^j = \theta ^0 + \sum _j \Theta^j $ where $\theta^0$ is an initial phase and
\ba
\label{Theta}
\Theta^j = \int _{x_+^{j-1}}^{x_-^j } \omega(z) \, \mathrm{d}z \, .
\ea
The key difference in our case compared to the case of $m^2\phi^2$ theory in flat background
studied in \cite{Dufaux:2006ee} is that, because of the  non-periodicity of $\omega^2(x)$,
The local maxima and minima of $\omega^2(x)$ increases linearly with $j$.
This results in a non-trivial $j$-dependence in $X^j$ and $\Theta^j$. Consequently, the occupation number of the $\hat \chi$ particles after $j$ oscillations of the inflaton is
\ba
\label{n}
n_{\K=0}^{j} = \vert\beta^{j}\vert ^2 = \exp \left(\sum_{\ell=1}^{j} 2X^\ell \right) \prod_{s=1}^{j}( 2 \cos \Theta ^s ) \, .
\ea

In order to provide some useful analytical expression for $n_{\K=0}^{j}$, we need to perform some reasonable approximations in calculating the integral for $X^j$ and $\Theta^j$
in (\ref{Xj}) and (\ref{Theta}). To calculate the integral in $X^j$  we note that the bounds of integration in (\ref{Xj}) ranges from $x^{min}_{j}-T/4$ to
$x^{min}_{j} +T/4$. This suggests that as a good approximation, we can approximate
$\Omega(x) = p x f(x) \simeq p x_{min} f(x)$ and the integral in  (\ref{Xj}) is
approximately  (for the details see {\bf Appendix \ref{approximation}})
\ba
\label{X-int}
X^j = \int ^ {x^{min}_{j} +T/4} _{x^{min}_{j}-T/4} \Omega(x) \, \mathrm{d}x \simeq \sqrt{p x^{min}_{j}}\int _{-\frac{T}{4}}^{+\frac{T}{4}} \sqrt{f(x)} \, \mathrm{d}x = \sqrt{p\pi K(1/2)} \,\dfrac{\Gamma \left(\frac{3}{8}\right)}{\Gamma \left(\frac{7}{8}\right)} \sqrt{2j - 1} \, ,
\ea
where $\Gamma(x)$ is the Gamma function, and $K$ is the complete elliptic integral of the first kind \cite{abramowitz}.
Similarly, for $\Theta^j$
 \ba
\Theta^j = \int ^ {x^{max}_{j} +T/4} _{x^{max}_{j}-T/4} \omega(x) \, \mathrm{d}x \simeq \sqrt{p x^{max}_{j}}\int _{-\frac{T}{4}}^{+\frac{T}{4}} \sqrt{f(x)} \, \mathrm{d}x = \sqrt{p\pi K(1/2)} \,\dfrac{\Gamma \left(\frac{3}{8}\right)}{\Gamma \left(\frac{7}{8}\right)} \sqrt{2j - 2} \, .
\ea
Using the approximations proposed for harmonic numbers summation in {\bf Appendix  \ref{harmonicnumbers}} one finds $\sum_{i=1}^j \sqrt{2i -1} \simeq 2 \sqrt2 j^{3/2}/3$ and
\ba
\label{X-sum}
\sum_1^{j } 2X^j  \simeq \dfrac{4}{3}\sqrt{2p\pi K(1/2)} \,\dfrac{\Gamma \left(\frac{3}{8}\right)}{\Gamma \left(\frac{7}{8}\right)} \,\,j^{3/2}  \, .
\ea
After $ j$ oscillations, we have $x= j T $, so for large $ j$ the exponent  for
$n_{\K=0}$ in (\ref{n}) behaves as
\begin{equation}
\label{n1k=0}
n_{\K=0}^j \propto \exp(0.490 \sqrt{p} \, x^{3/2} )  \, .
\end{equation}
Quite interestingly, the occupation number grows exponentially with the exponent proportional to $x^{3/2}$. One can trace the non-linear exponential enhancement of the occupation number to the explicit breaking of the periodicity
of $\omega^2(x)$ in (\ref{omega}). This is in contrast to conventional model of preheating where the source term is periodic and the Floquet theorem applies with the result that  the occupation number has a linear exponential growth.
One can  check that the breaking of periodicity in our case
is a direct consequence of conformal invariance breaking
via trilinear interaction in an expanding background.

\begin{figure}[t]
\includegraphics[scale=.25]{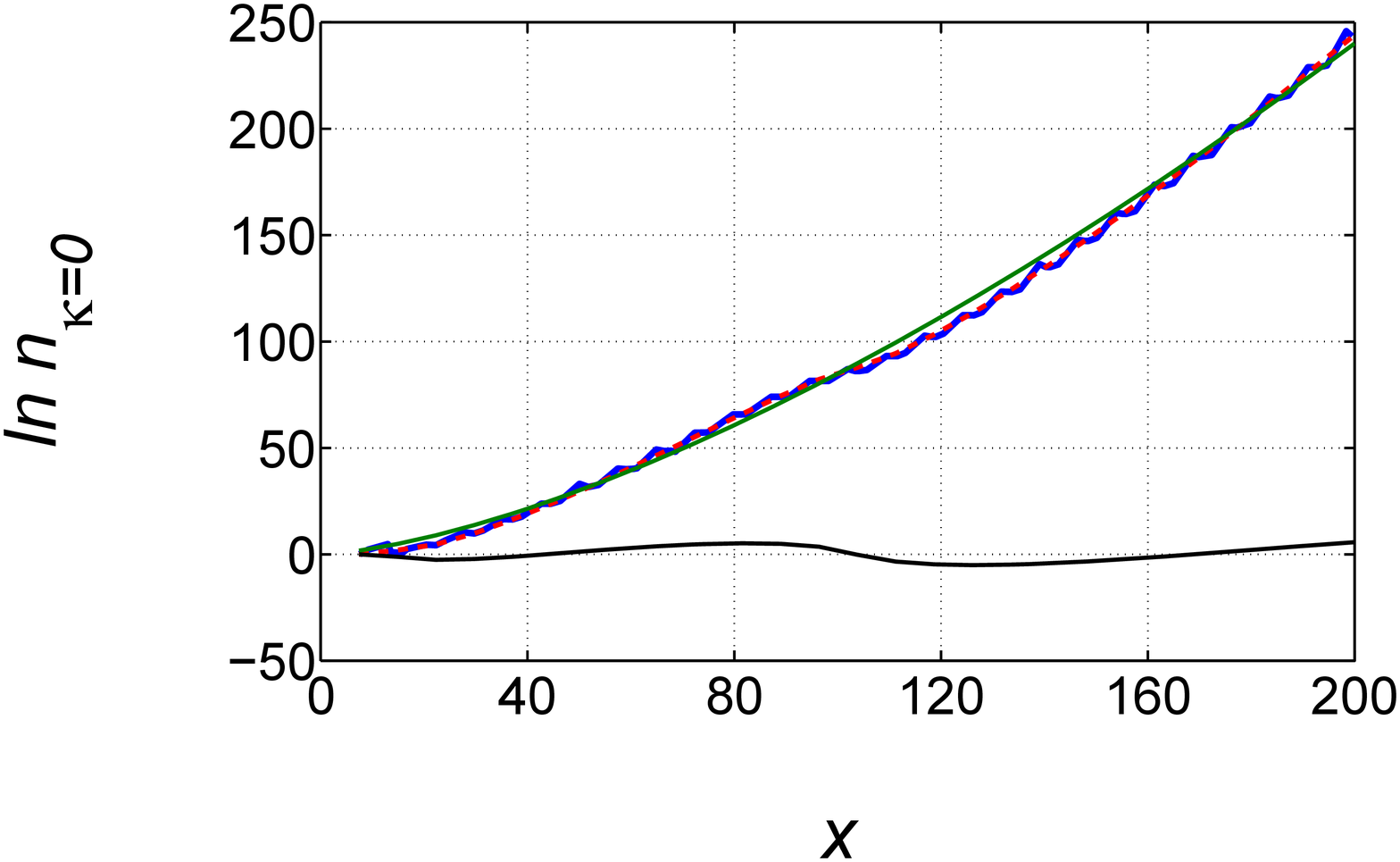}
\includegraphics[scale=.25]{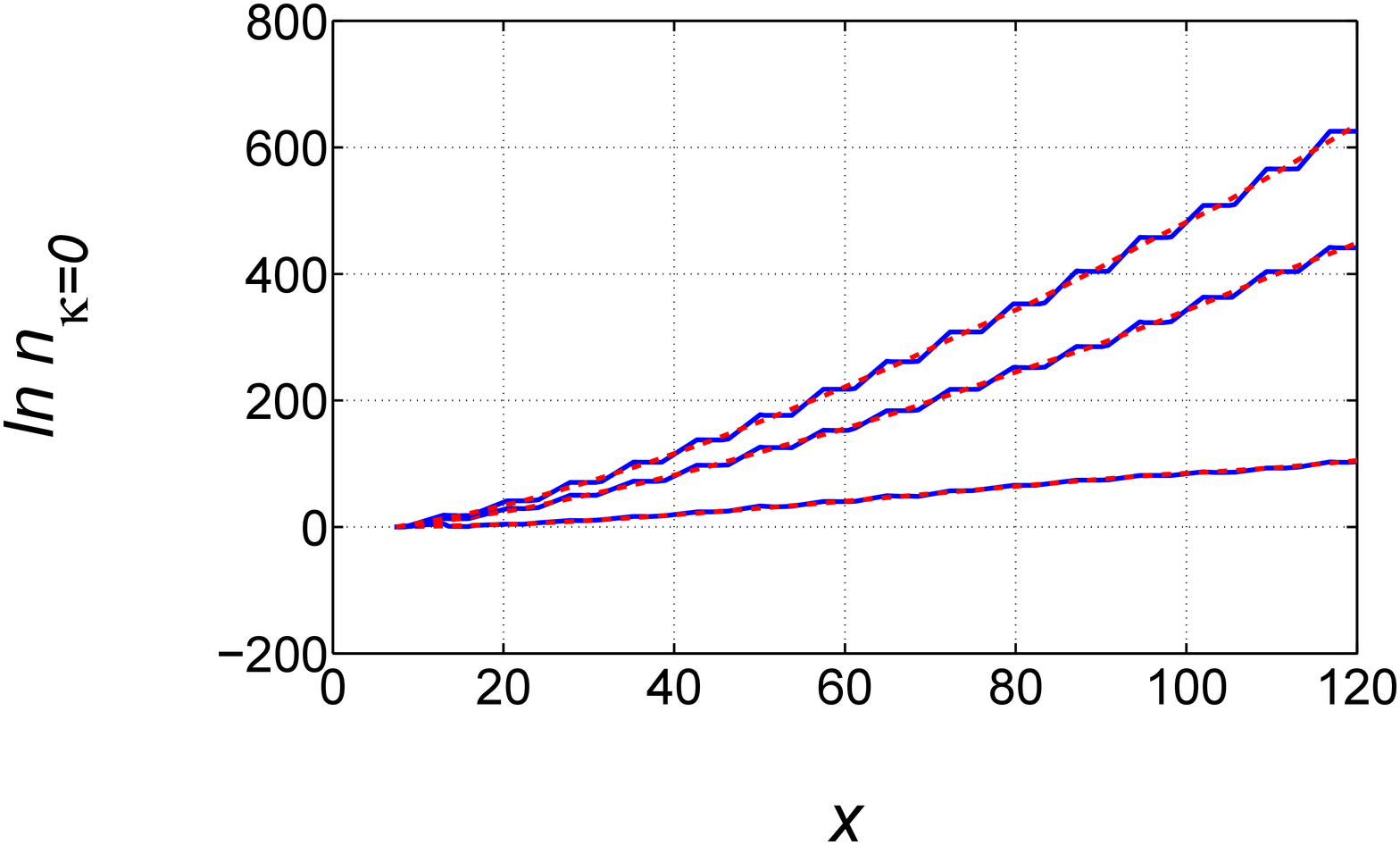}
\caption{\label{p010305}
Occupation number $\ln n_{\K=0} $ as a function of conformal time $x$.
In the left figure with $p=0.03$,  the wavy solid curve (blue)  shows the exact numerical solution, the smooth growing curve (green) shows the analytic solution (\ref{n1k=0}) which does not contain the interference term, the  dash-dotted curve (red) shows the analytic solution (\ref{n2k=0}) which includes the  interference term added and the bottom solid curve (black) shows the behavior of the interference term. In the right figure, $\ln n_{\K=0} $ is plotted for different values of $p$. Again the solid curves (blue) show the exact numerical solutions of  (\ref{X-eq}) while the dashed curves (red) shows the analytic solution (\ref{n2k=0}).  The graphs from bottom to top, respectively, correspond to $p=0.03, p= 0.5$ and $p=1$. The agreement between the full numerical results and our analytical formula, (\ref{n2k=0}), is impressive.  }
\end{figure}

A plot of $n_{\K=0}$ is shown in the left graph of {\bf Fig. \ref{p010305}} containing the full numerical solutions and our analytical results. Although the non-linear
exponential growth captures very well the overall behavior of $n_{\K=0}$, but the figure
from the exact numerical solution shows some small wiggles and ripples
which can not be captured by the
exponential profile.  To take  these minor but interesting discrepancies
into account, we should also add the effects of the phase term originated
from the products of cosines in (\ref{n}).
As one can see in {\bf Fig. \ref{p010305} } this term is oscillatory
but not periodic which can explain the small oscillations in the profile
of  $\mathrm{ln}\, n_{\K=0} $ versus $x$.  Taking
the effect of the phase term into account, one obtains
\begin{equation}
\label{n2k=0}
n_{\K=0}^j \simeq \exp\left[0.490 \sqrt{p} \, x^{3/2} + \sum_{\ell=1}^{j} \ln \left( 4 \cos^2 \left( 7.427 \sqrt{p\,  \ell} \right) \right) \right] \, ,
\end{equation}
where $j=x/T$.

In the right graph of {\bf Fig. \ref{p010305}} we have plotted $n_{\K=0}$ for different values
of $p$. As this graph shows, with the addition of the effects of the phases, our analytical result (\ref{n2k=0})
shows a perfect agreement with the
exact numerical results, confirming that our analytical solution captures the correct $p$-dependence.
As one expects, the larger  the value of $p$ is, the stronger is  tachyonic resonance from  trilinear interaction.

\subsection{ Tachyonic Resonance with  $\kappa^2 \neq 0$}
\label{non-zero-k}

After presenting the analysis for the simple  zero-momentum case, in this section we present
the analysis for arbitrary momentum, $\kappa^2 \neq 0$. As can be seen from
{\bf Fig. \ref{rfx}}, for $\kappa^2 \neq 0$ there is no tachyonic resonance from the beginning
where $\kappa^2 > p\, x$.  After some oscillations $\omega^2(x)$ becomes negative for $x\ge p/\kappa^2$ and our previous methods for tachyonic resonance would apply.
From \eqref{omega}, we see that the frequency of oscillations becomes tachyonic after $ j = j _{\ast}$ oscillations where
\begin{equation}
\label{j-star}
j_{\ast} = [ \dfrac{\kappa^2}{p T } + \dfrac{1}{2}] \, .
\end{equation}
Here $[z]$ represents the integer part of $z$.

For small $x$ there is no tachyonic region from the start of preheating, but if the frequency becomes non-adiabatic one should expect particle production via parametric resonance
as in conventional preheating analysis. However, as we shall see from our full numerical results, the particle creation from the non-tachyonic scattering for $j< j_*$ oscillations is quite negligible compared to tachyonic resonance particle creation after $j>j_*$ oscillations.
The non-adiabadicity condition  $ |\omega'(x)|/\omega(x)^2 \gg1$
is satisfied for
\begin{equation}
j \geq j_{\mathrm{nad}} \simeq j_{\ast} - \dfrac{0.8472}{\pi \sqrt{p T}} \sqrt{j_{\ast}}\simeq j_{\ast}-\frac{0.1}{\sqrt{p}}\ \sqrt{j_{\ast}}  \, .
\end{equation}
This indicates that $j_{\mathrm{nad}} - j_* \sim$ few and the onset of parametric resonance is actually around $j_*$ where tachyonic resonance starts. Besides the very short period of parametric resonance,
the ``effective'' Floquet index during this period is much smaller than that of the tachyonic regions.
Therefore,  the particle production is mainly concentrated in the tachyonic regions. This result is also verified in our numerical investigations.

Repeating the same methods as in the previous section,  for $j> j_*$,
the Bogoliubov coefficients before and after $j^{th}$ oscillations are related by
\begin{equation}
\begin{pmatrix}
\alpha_\K^{j + 1} \\
\beta_\K^{j +1}
\end{pmatrix}
= e^{X_\K^j}
\begin{pmatrix}
1 & i e^{2i\theta^j} \\
-i e^{-2i\theta^j} & 1
\end{pmatrix}
\begin{pmatrix}
\alpha_\K^{j } \\
\beta_\K^{j }\ ,
\end{pmatrix}
\end{equation}
where
\begin{equation}
\label{X-int2}
X_\K^j  \simeq \sqrt{p\,  x_{j}^{min}} \int_{x_{j}^{-}}^{x_{j}^{+}} \sqrt{\vert r_j^{min} + f(z) \vert } \mathrm{d} z \, ,
\end{equation}
with $r_{j}^{min} \equiv \kappa^2/p x_j^{min}$
and $\theta^j $ is the total phase accumulated from $x_0$ to $x^-_j$ during the intervals where $\omega^2 > 0 $. We have $\theta^j = \theta ^0 + \sum _j \Theta^j $ with
\begin{equation}
\label{Theta-int2}
\Theta^j \simeq \sqrt{p \, x_{j}^{max}} \int_{x_{j-1}^{+}}^{x_{j}^{-}} \sqrt{r_j^{max} + f(z)} \mathrm{d} z  \, ,
\end{equation}
where  $r_{j}^{max} \equiv \kappa^2/p x_j^{max}$.

Similar to the analysis  resulting in (\ref{X-sum}) and using the formulas presented in
{\bf Appendix \ref{r+jacobicn}}, one obtains
\ba
\label{X-sum2}
\sum_{j_{\ast}+1}^{j} 2 X_{\K}^{\ell} &\simeq&  \sum_{\ell=j_{\ast}+1}^{j} \left( 2a \sqrt{\dfrac{pT}{2} } \sqrt{2\ell-1} - 2b'\kappa^2 \sqrt{\dfrac{2}{pT} } \dfrac{1}{\sqrt{2\ell-1}} \right) \nonumber\\
&\simeq&\dfrac{4a}{3} \sqrt{pT} (j^{3/2}-j_{\ast}^{3/2}) - 4 b'\kappa^2 \sqrt{\dfrac{1}{pT} } (j^{1/2}-j_{\ast}^{1/2})\\
\label{phaseterm}
\prod_{j_{\ast}+1}^{j}( 2 \cos \Theta_\K ^j )^2 &=& \exp \left[\sum_{j_{\ast}+1}^{j} \ln \left( 4 \cos^2 \left( a \sqrt{pT} \sqrt{j - 1} + \dfrac{b \kappa^2}{\sqrt{pT}} \dfrac{1}{\sqrt{j-1}}\right) \right) \right] \, ,
\ea
where \eqref{1/j} and \eqref{2j-1} have been used and
\ba\label{a,b}%
a=2.72,\qquad b'=2.86,\qquad b=3.75\ .
\ea%
\begin{figure}[!t]
\includegraphics[scale=.25]{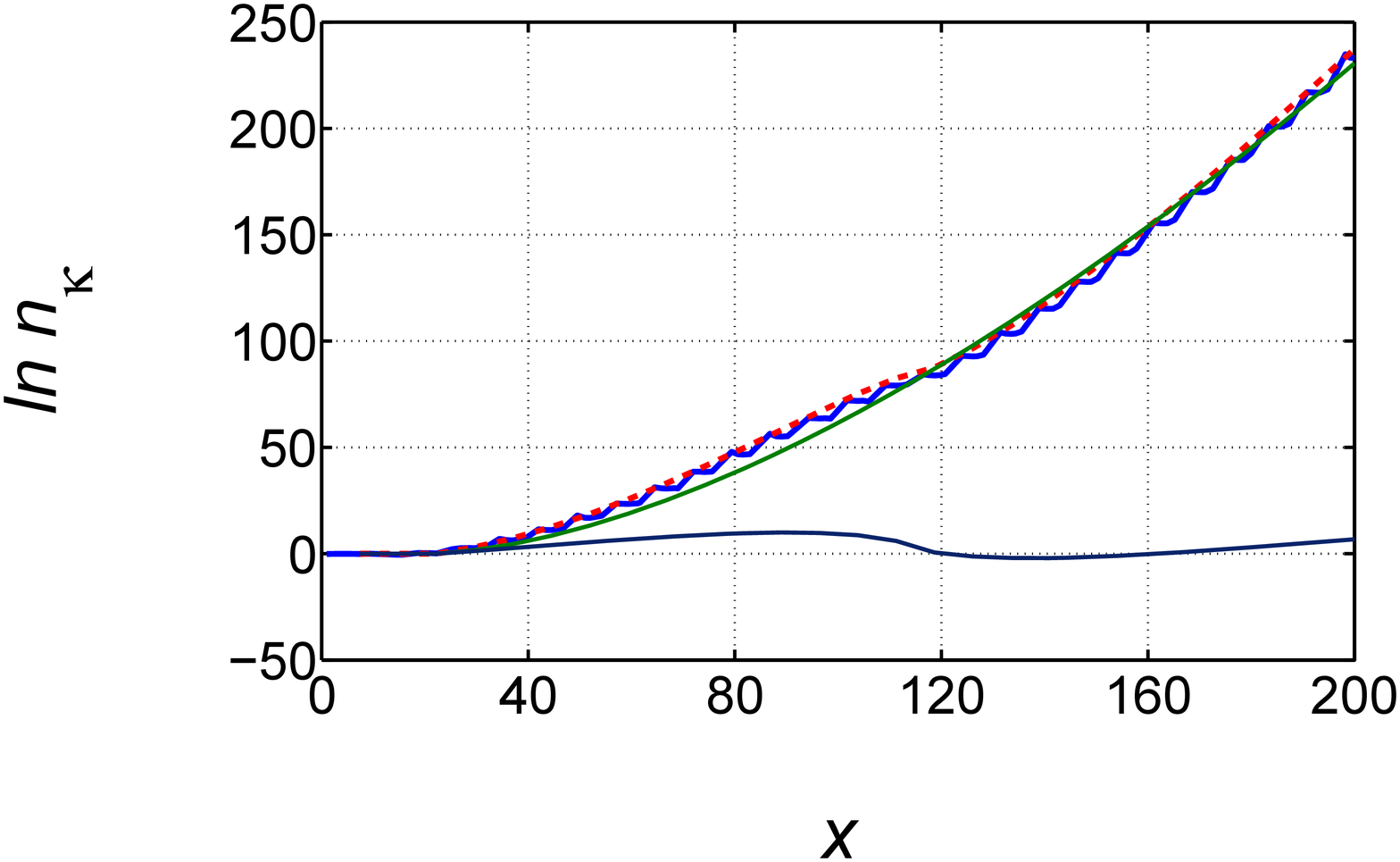}
\includegraphics[scale=.25]{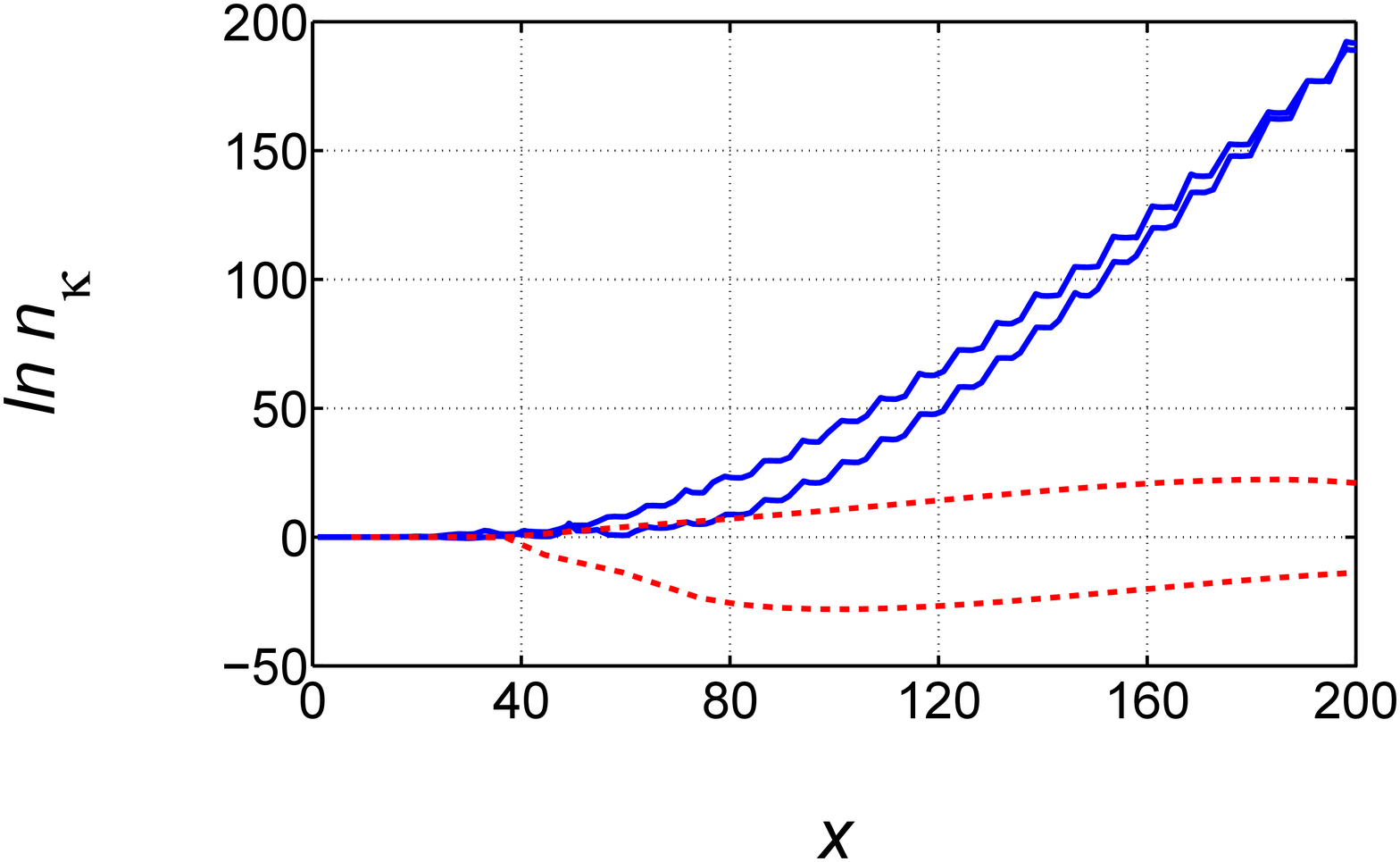}
\caption{\label{comparison}
The left figure shows $\ln n_{\K}$ as a function of $x$ for $p=0.05$ and $\kappa^2=1$.
The wavy solid (blue) curve shows the exact numerical solution of \eqref{X-eq}. The (red) dashed curve shows the  analytical solution (\ref{nwithphase}).
The smooth (green) solid curve shows the analytical solution
without taking into account the interference term. The bottom (black) solid curve shows the effect of the interference term (\ref{phaseterm}).
The agreement between the full numerical result and the analytical formula (\ref{nwithphase}) is impressive. In the right figure shows the destructive effects of the interference term
in (\ref{nwithphase}). The  lower and upper wavy (blue) solid curves show the exact numerical results for $\kappa^2 = 2$ and $\kappa^2 = 1.5$, respectively, with $p=0.05$.
The lower and upper dashed (red)
curves show the effects of the interference term, for  $\kappa^2 = 2$ and $\kappa^2 = 1.5$, respectively. We see that the interference term causes a suppression of particle production
for $\kappa^2 = 1.5$. }
\end{figure}
After $j^{th}$ oscillations, $x = jT$, and  the occupation number is given by
\begin{equation}
\label{nwithphase}
n_\K^j \sim \exp \left( \dfrac{4a \sqrt{p}}{3T} \, \left(x^{3/2} - x_{\ast}^{3/2} \right) - \dfrac{4b'\kappa^2}{\sqrt{p}T} \left( x^{1/2} -x_{\ast}^{1/2} \right) \right)
\times \mathrm{Interference ~ term}  \, ,
\end{equation}
where $x_* \equiv j_* T$ and the interference term is the expression \eqref{phaseterm}.

As in the zero-momentum example, the occupation number has a non-linear exponential dependence with the leading exponents being $x^{3/2}$ and $x^{1/2}$ respectively. The fact that before the tachyonic regions, i.e. for $x<x_*$, there is no particle creation is clearly seen in (\ref{nwithphase}). Both of these two non-trivial results are a consequence of the violation of the periodicity in the resonance source term. The left plot
in {\bf Fig. \ref{comparison}} compares our analytical result (\ref{nwithphase})
with the full numerical results. The agreement is  again impressive.

 The interference term in (\ref{nwithphase}) has a very interesting effect.
Suppose for the moment that we do not take into account the interference term in (\ref{nwithphase}). Since $b'>0$,
from the exponential term one may expect that the larger the value of $\kappa^2$,
the more suppressed will the particle creation be. Our numerical
investigations supports this general rule, but there are some noticeable
exceptions. In the right graph of {\bf Fig. \ref{comparison}} we have  presented an example where a larger $\kappa^2$ has a higher occupation number than
a smaller $\kappa^2$. The resolution to this apparent paradox relies on the destructive effects of the the interference term. Looking into the form of the interference term (\ref{phaseterm}) one observes that the interference becomes destructive in a region
in $\kappa^2-j$ space where $\cos(\Theta_\K^j)$ vanishes. For the latter, one requires  the
phase term $\Theta_\K^j$ to be stationary  on its roots, i.e. $\cos(\Theta_\K^{j}) = \partial_j \Theta_\K^j=0$ at some stationary time $j_s$. From  (\ref{phaseterm}) one easily finds
that the stationary points of $\Theta_\K^j$ occur at
\ba
\label{js}
j_s= 1+ \frac{b \kappa^2}{a p T} \, ,
\ea
with
\ba
\Theta_\K^{j_s}= 2 \sqrt{ab} \kappa \, .
\ea

As explained above, demanding that  $\Theta_\K^{j_s} \rightarrow
(2n-1)\frac{\pi}{2}$ for some integer $n$, one obtains the value of
$\kappa$ where the destructive feature of the interference term is
pronounced \ba \label{kappa-stab} \kappa_{\mathrm{stab}} =
\dfrac{(2n-1)\pi}{4\sqrt{ab}}\simeq 0.25 (2n-1)  \ , \quad
n=1,2.... \ea

Note that $\Theta_\K^{j_s}$ and $\kappa_{\mathrm{stab}}$ are
independent of $p$. Note also that  the stable bands are equally-spaced  in the
momentum space. It is worth checking this equation for some $n$.
For example for $n= 2,\,3,\ 4 \,\mathrm{and} \,5$, one finds $\kappa^2 = 0.54
,\, 1.51 ,\, 2.96\, \mathrm{and}\, 4.89$ respectively.
\begin{figure}[t]
\includegraphics[scale=.45]{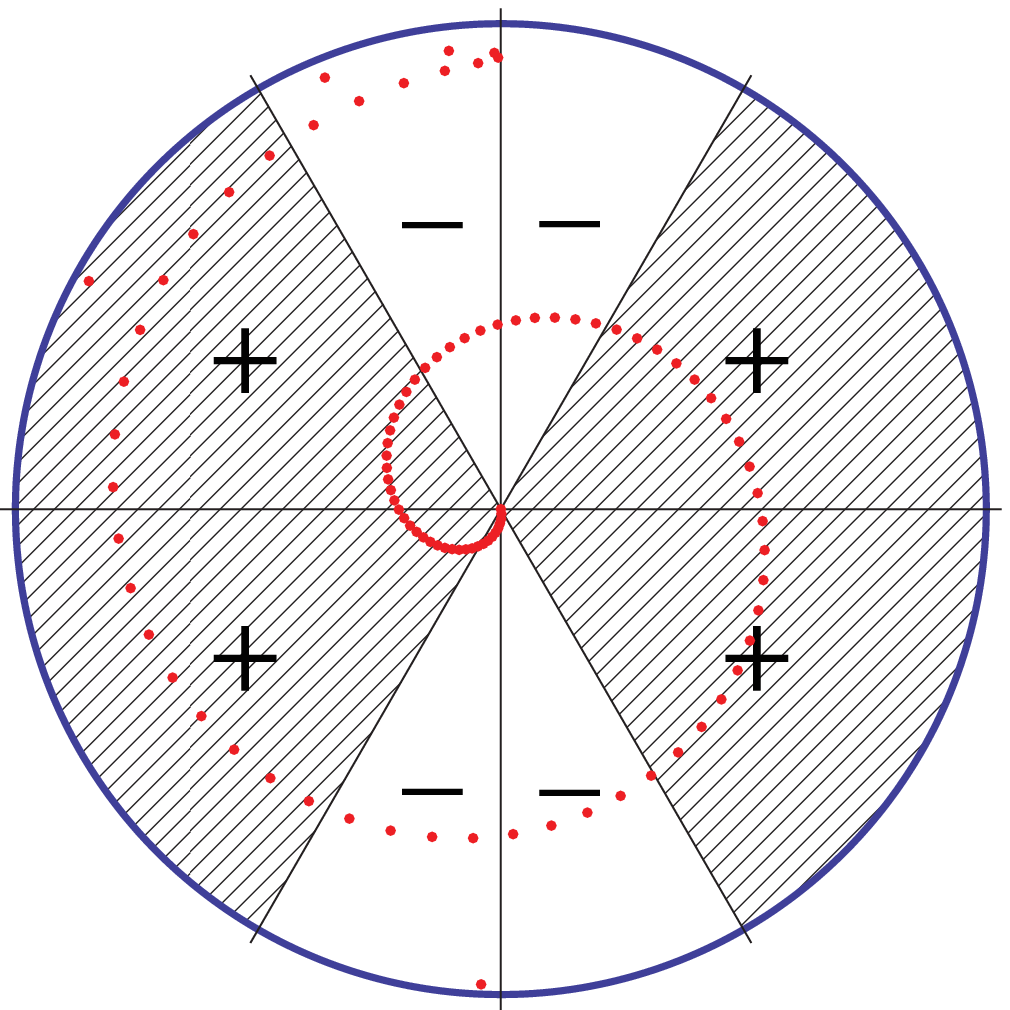}\hspace{1cm}
\includegraphics[scale=.45]{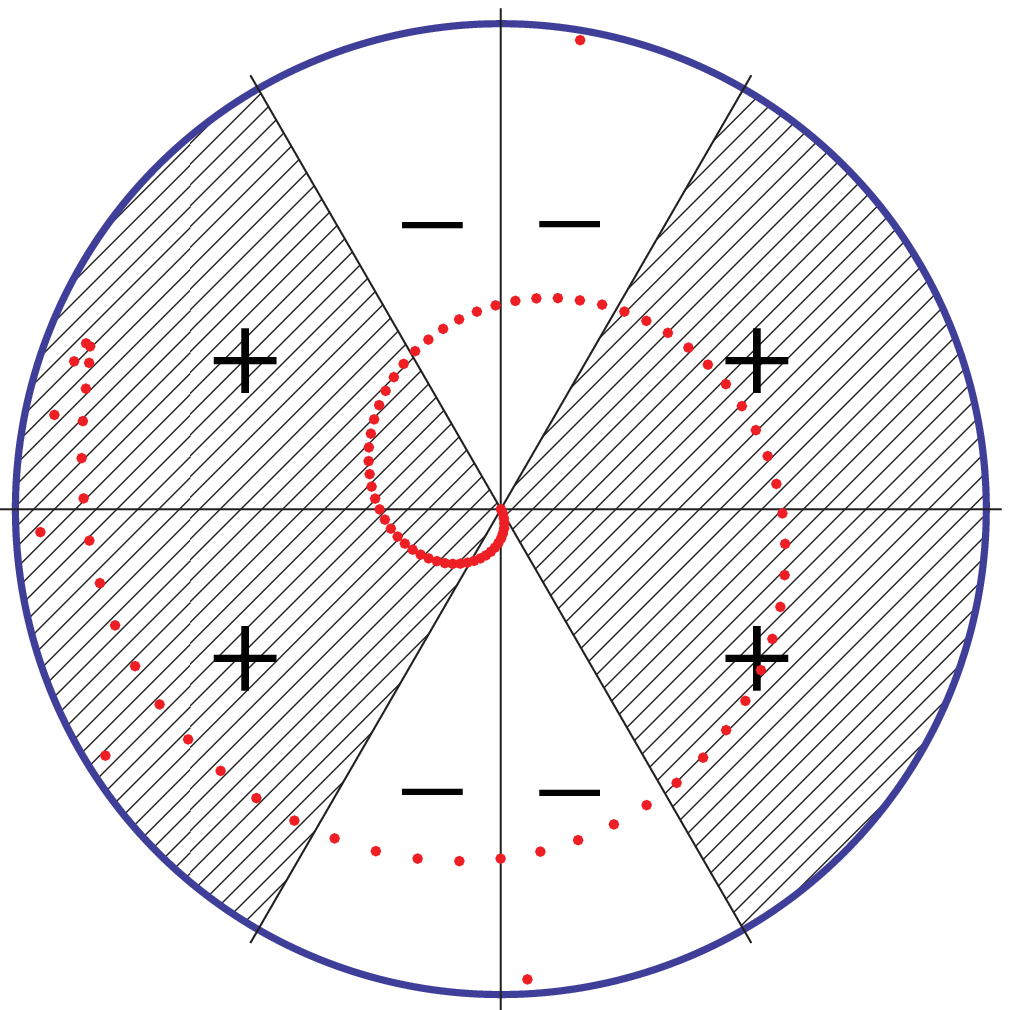}\hspace{1cm}
\includegraphics[scale=.45]{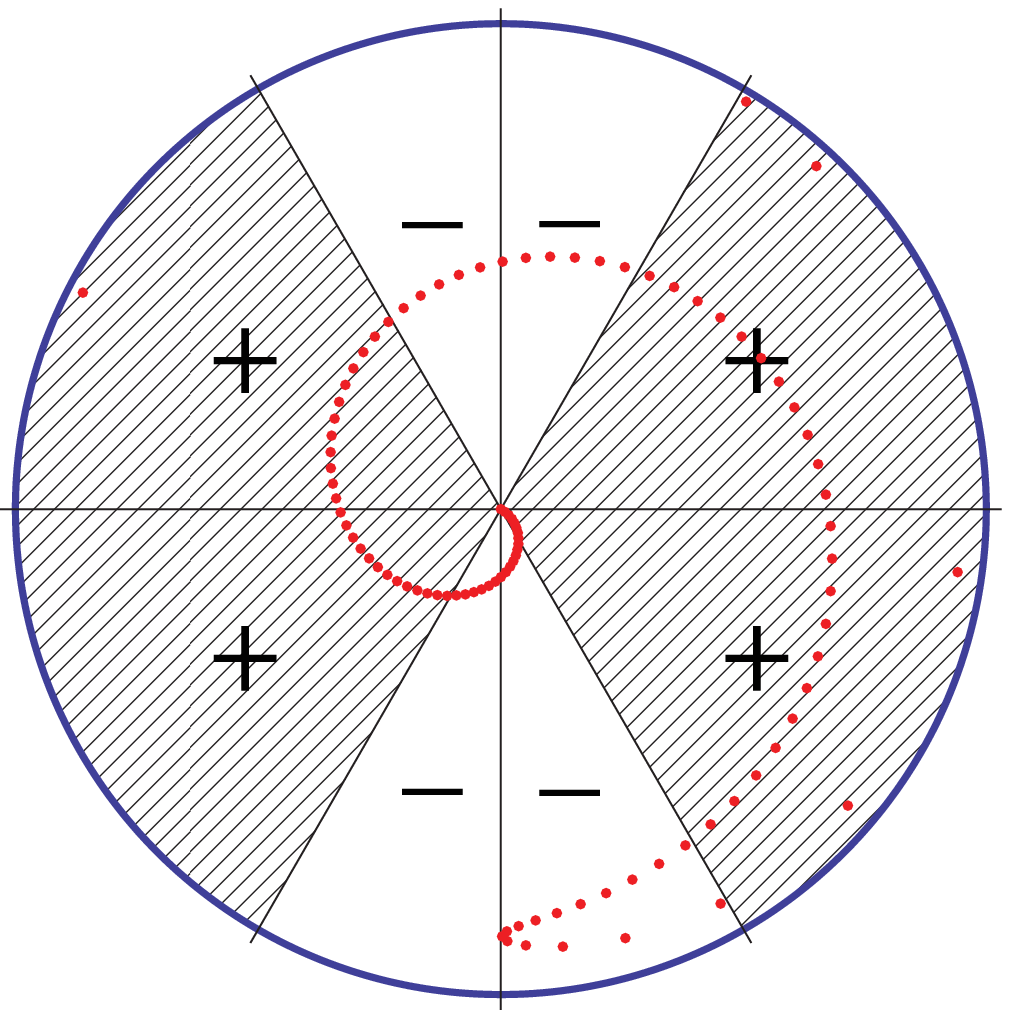}
\caption{\label{dayere}
Phase accumulated  through $j$ oscillations for different values of $\kappa^2$,
left: $\kappa^2=1.51$, middle:  $\kappa^2=2$ and  right: $\kappa^2=2.96$.
The angular position of each (red) dot indicates the value of $\Theta_\K^j$. The outer points correspond to smaller value of $j$ and as $j$ increases the dots become denser as can be seen from
the form of $\Theta_\K^j$ given in (\ref{phaseterm}).
As $\Theta_k^j$ tends to $(2n-1)\pi/2$, the contribution of $j^{th}$ oscillation in the interference-term suppresses the  particle production. For $\kappa=1.51$ (left) the stationary point of $\Theta_k^j$ is near $\pi/2$ . For $\kappa=2$ (middle) the stationary point is far away
from the poles and for $\kappa=2.96$ (right) the stationary point is near $3\pi/2$.
This explains why $\kappa^2=1.51$ and $\kappa^2=2.96$ belong to the ``semi-stability bands''.
Note that $\ln | 2 \cos \Theta_\K^j | $ is positive for $|\Theta_k^j - n\pi|<\pi/3$ and is negative for $|\Theta_k^j - (2n+1)\pi/2|<\pi/6$, which are separated  in this figure by $+$ and $-$ signs. }
\end{figure}
{\bf Fig.\ref{dayere}} shows behavior of $\Theta_\K^j$ as a function of
number of oscillations for three given values of $\kappa^2$. As is seen in the left and right figures the stationary point of $\Theta_\K^{j_s}$ can be near $(2n-1)\pi/2$ which causes particle production suppression.

One also observes that the destructive phase in the interference term is not static in time.
This is somewhat similar to stochastic resonance phenomena observed in resonance preheating in an expanding background \cite{Kofman:1997yn}.  However, in contrast to the stochastic resonance, in our case the interference term is not completely stochastic and one can keep track of the phase at each oscillation period. The value of $\kappa$ around $\kappa_{\mathrm{stab}}$, where the destructive effects of the interference term is significant, form \textit{semi stability bands}.
These semi stability bands, as indicated in \eqref{js}, occur at
\ba\label{x-stability}%
x_{\mathrm{stab}}=T+\frac{\pi^2}{16a^2\ p}\ (2n-1)^2\simeq 7.416+0.0834 \frac{(2n-1)^2}{p}\ .
\ea%

\begin{figure}[t]
\includegraphics[scale=.25]{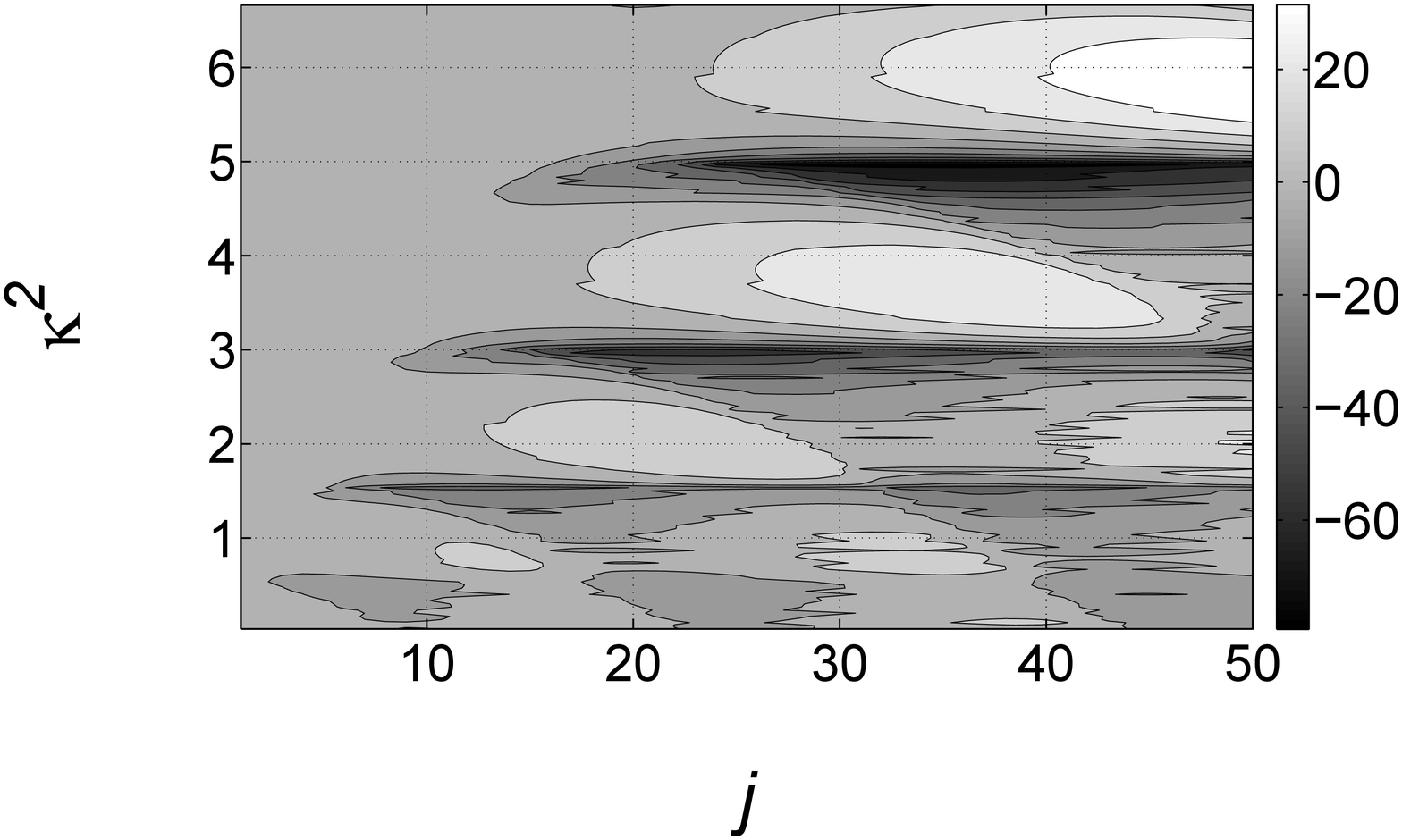}
\includegraphics[scale=.25]{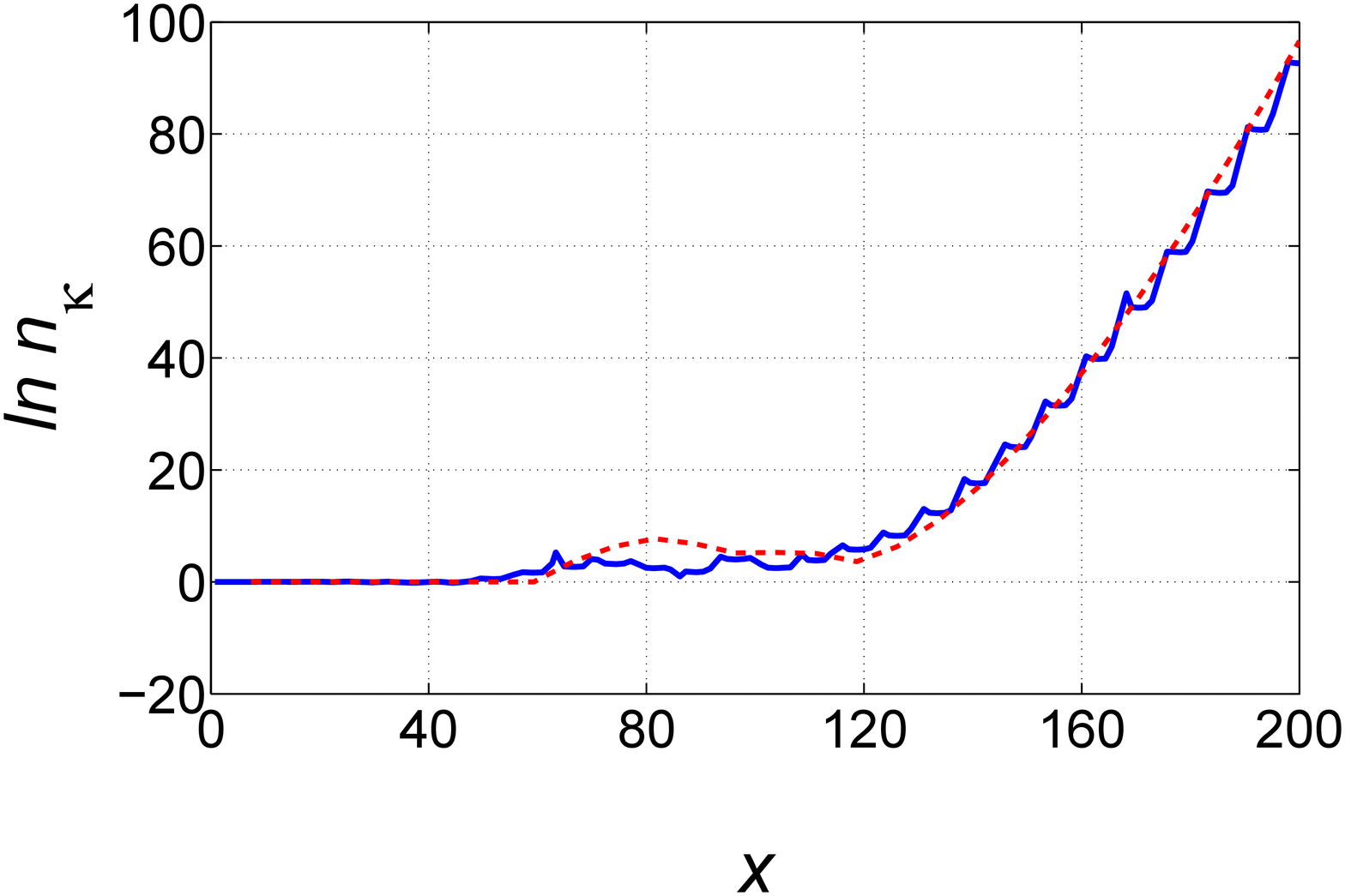}
\caption{\label{phasecontour}
In the left figure the contour plot of the phase term in $\kappa^2 - j$ plane is shown.
The darker regions show the semi-stability bands in which the interference term suppresses particle creation. The right bar represents $\ln |2 \cos \Theta^j |$
as a function of number of oscillations $j$. An important point of our analysis is that the semi-stability regions are independent of $p$. In the right figure, $\ln n_{\K}$ as a function of conformal time for the predicted semi-stability band $\kappa^2\simeq 3$  is shown. The
(blue) solid line shows the full numerical solutions and the (red) dashed curve shows our analytical
 result (\ref{nwithphase}). As we predicted,   particle production for this momentum  is suppressed for small times because of the interference term.  }
\end{figure}

A plot of the effects of the interference terms is shown in {\bf Fig.\ref{phasecontour}} with $p=0.05$.  As an example of the evolution of the semi-stability band consider the line
$\kappa^2=1.5$ which corresponds to $n=3$ in  (\ref{kappa-stab}). For small $x$,
the interference term has a destructive contribution. As the time goes by,
effects of the interference term would become milder and less important.
As explained above, for this reason we may call $\kappa^2=1.5$ a
semi-stable band rather than a stable band. As shown above, the other
semi-stable band is at $\kappa^2 \simeq 2.96$. The tachyonic resonance for
this case would start at $x_*= j_* T \simeq 60 $. Nonetheless, as depicted
in the right graph in {\bf Fig.\ref{phasecontour}} particle creation is negligible until $x \simeq 110$; compatible with $x_\mathrm{stab}\simeq 104$ obtained from \eqref{x-stability} for $n=4$.
This is a consequence of
the semi-stability band for $\kappa^2 = 2.96$ which suppresses the
particle creation at early time and eventually loses its importance as
the Universe expands further.

\subsection{Expanding vs. non-expanding background}

In the last two subsections the effects of expanding background were included in the preheating analysis via the conformal transformations $\varphi =  a(\eta) \phi$ and $\hat \chi= a(\eta) \chi$. One consequence of the expanding background was the appearance of the non-periodic factor $x$ in $\omega(x)$. This in turn leads to a non-linear exponential particle creation. This should be compared with preheating with four legs interaction, $g^2 \phi^2 \chi^2$, in $\lambda \phi^4/4$ theory which preserves the conformal invariance and $\omega^2(x)$ is periodic and one obtains a linear exponential
particle creation.  One may ask what would be the situation if one considers preheating with trilinear interaction in a non-expanding flat background.  The trilinear term in a flat background
is periodic and, as a consequence of the Floquet theorem, one obtains a linear exponential particle creation. This seems paradoxical,  noting that the expansion
of the Universe usually suppresses the particle creation. As we show below, the root of this apparent  paradox relies on difference between conformal time $x$ and the cosmic comoving time $t$.

\begin{figure}[t]
\includegraphics[scale=.35]{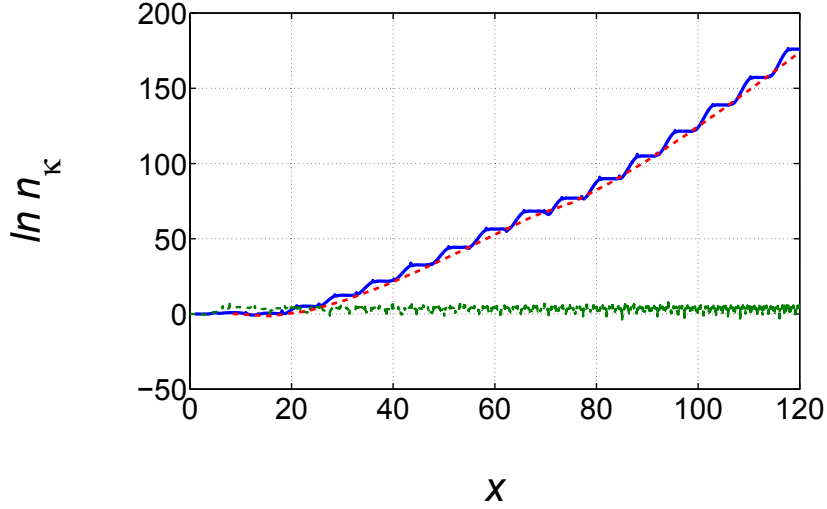}
\caption{ Comparison between particle creations in a flat  and
expanding backgrounds in $\lambda \phi^4/4$ theory. For $p= 0.1$ and $\kappa^2=0.54$ which is in
tachyonic region in a flat background,
but because of effect of interference term ({\ref{XTnonexp}}) there is
no particle production for these parameters. Nonetheless, due to  non-static
interference-term, there is particle production in expanding universe.
The wavy (blue) solid curve indicates the full numerical solution for an
expanding Universe, the (red) dashed curve shows our analytical solution
and the straight (green) line at the bottom represents the particle creation in a flat background.
\label{phi4-noexpansion}
 }
\end{figure}

In a flat background with  $a(t)=1$, the solution to the inflaton field is $\phi = \tilde \phi f(z)$ where $\tilde \phi$ is the initial amplitude of the inflaton field at the start of preheating and $f(z)$ is the Jacobi elliptic cosine function. Here we defined the dimensionless times $z\equiv \sqrt \lambda \tilde \phi t$. The equation of motion for the resonant field $\chi$ is
\ba
\label{X-eq-z-coordinate}
\frac{d^2 \chi_\K}{d z^2} + \left( \kappa^2 + p x_0 f(z) \right)\chi_\K  =0
\ea
 with $x_0 = \sqrt{3/ 2\pi} (\mpl / \tilde{\varphi})$ and $\kappa^2 = k^2/\lambda \tilde \varphi^2$.

 Following the same methods as in the last two subsections, the occupation number is
\begin{equation}
n_k^j = \exp (2jX_k)(2 \cos\Theta_k)^{2(j-1)}
\end{equation}
with the crucial difference that now $X_k$ is $j$-independent as in \cite{Dufaux:2006ee}. To a good approximation, one has
\begin{equation}
\label{XTnonexp}
X_k \simeq  a\sqrt{p x_0}-\dfrac{b'}{\sqrt{p x_0}}\kappa^2, \qquad \mathrm{and} \qquad \Theta_k =  a\sqrt{p x_0}+\dfrac{b}{\sqrt{p x_0}}\kappa^2\ ,
\end{equation}
where the numeric coefficients $a,b,b'$ have the same values as before, given in \eqref{a,b}.
By substituting $j \rightarrow \tau/T$, where $T$ is the period of
oscillations of $f(z)$ \eqref{periodT}, the  occupation number of the $\K$-mode becomes
\begin{equation}
\label{nk-non-expanding}
n_\K^j = \exp \left( {\dfrac{2 X_\K \sqrt \lambda \tilde \phi}{T} t} \right)  \times  \mathrm{interference\,\, term}\ .
\end{equation}

As expected $n_\K$ has a linear exponential growth in terms of $ t$. To compare it with the particle creation in an expanding background, we recall that in our case  $t \propto x^2$ (\textit{cf.} \eqref{x-def}) and hence
(\ref{nk-non-expanding}) gives $\ln n_\K \sim x^2$ for the flat background. This confirms the intuition that in general  the particle creation via  tachyonic resonance is more enhanced in a flat background compared to that of an expanding background.
This is understandable because the  expansion of the Universe dilutes the previously produced particles and also reduces the amplitude of the source term $\phi(t)$.  One should, however, note that due to the non-trivial dynamics of the interference term, this intuition may not work and for some specific regions in the $\kappa$ space expansion of the Universe may  enhance the particle production rate.
As explained in previous subsection, the interference term is time dependent which means the stable or unstable bands vary in time. As an example,  {\bf Fig.\ref{phi4-noexpansion}} shows cases in which the particle production is more efficient in an expanding background.





\section{Tachyonic resonance in $m^2\phi^2/2$ theory}
\label{m2phi2-case}
The $\lambda \phi^4/4$ inflationary model has the conformal symmetry and the effects of
the expanding background can be incorporated by the conformal transformations $(\phi, \chi) \rightarrow a(\eta) (\phi, \chi)$. This trick does not apply to $m^2 \phi^2/2$ model and the effect of expansion should be taken care of accordingly. The analysis of the tachyonic resonance for $m^2 \phi^2/2$ model
in a flat background has been studied in \cite{Dufaux:2006ee} with some brief discussions on the effects of expanding background. Here we study the tachyonic resonance for $m^2 \phi^2/2$ inflationary potential in an expanding background in more details and demonstrate that it can have very nontrivial consequences.

We start with the potential
\begin{equation}\label{m2phi2-potential}
V(\phi) = \dfrac{m^2}{2} \phi^2 + \dfrac{\sigma}{2}\phi \chi^2  +
\dfrac{\lambda}{4} \chi ^4 \ ,
\end{equation}
with $\lambda > \sigma^2/2 m^2$ \cite{Dufaux:2006ee}.
The equation for the production of $\chi $ particles obeys
\begin{equation}
\label{mathieu}
\hat \chi''_{k} + (A_k + 2q \cos\,2z ) \hat \chi_{k} = 0\ ,
\end{equation}
where $\hat \chi = a(t)^{3/2} \chi$ and
\begin{equation}
mt \equiv  2z - \frac{\pi}{2},\qquad A_{k} = \frac{4k^2}{m^2 a^2}\equiv \frac{A^k_{0}}{a^2}\qquad \mathrm{and}\qquad q = \frac{2 \sigma \Phi_{0}}{m^2 a^{3/2}}\equiv\frac{q_{0}}{a^{3/2}} \, .
\end{equation}
Here prime denotes derivatives with respect to coordinate $z$ and
$A_0$ and $q_0$ indicate the values of the corresponding quantities in a flat background. For the quadratic potential, the background Universe during preheating
evolves like matter domination
and  $a(t)= a_0 (t/t_0)^{2/3}$ and in our conventions we choose $a_0 =1$ and $t_0=1$.

One can incorporate  effects of expansion in the tachyonic resonance analysis as follows.
The tachyonic regions of  (\ref{mathieu})  are centered around the minimum of
$\omega^2(z)$, at $t_j = (j-1/2)\pi$ with the scale factor $a_-(t_j)= ((j-1/2)\pi)^{2/3}$
whereas the non-tachyonic regions are centered
around the maximum of $\omega^2(z)$, at $t_j = (j-1)\pi$ with the scale factor $a_+(t_j)=((j-1)\pi)^{2/3} $.
Following the same steps
as in section  \ref{zero-k} the occupation number is given by
\begin{equation}
n_\K^{j} = \vert\beta_\K^{j}\vert ^2 = \exp\left(\sum_{j_{\ast}+1}^{j} 2X_\K^j \right) \prod_{j_{\ast}+1}^{j}( 2 \cos \Theta_\K ^j ) \, ,
\end{equation}
where $j_{\ast}$ is defined as the last non-tachyonic oscillation after which tachyonic resonance starts
\begin{equation}\label{j-star-m2phi2}
j_{\ast} = [ \dfrac{1}{\pi}\left(\dfrac{ A_k^0}{2q_0} \right)^3+\dfrac{1}{2}] \, ,
\end{equation}
where $[z]$ represents the integer part of $z$.

Using results of   \cite{Dufaux:2006ee}, one finds
\begin{equation}
X_{k}^{j} = - \dfrac{\alpha \,A^0_{k}}{\sqrt{q_{0}} a_-(t_j)^{5/4}} + 2 \alpha \dfrac{\sqrt{q_{0}}}{a_-(t_j)^{3/4}} = - \dfrac{\alpha\,A^0_{k}}{\sqrt{q_{0}} t_{j}^{5/6}} + 2 \alpha \dfrac{\sqrt{q_{0}}}{t_{j}^{1/2}} \, ,
\end{equation}
in which $\alpha = 0.85$
 and $t_{j} = \pi (j-1/2)$. Furthermore, the phase accumulation is given by
\ba
\label{phase-quadratic}
\Theta_{k}^{j} = \sqrt{\dfrac{ q_{0} }{ { a_+(t_j) }^{3/2} } }
\left[a + b \dfrac{A_{k}^{0}}{2q_{0} {a_+(t_j)}^{1/2}} + c \left(1- \dfrac{A_{k}^{0}}{2q_{0} {a_+(t_j)}^{1/2}} \right) \ln \left(1- \dfrac{A_{k}^{0}}{2q_{0} {a_+(t_j)}^{1/2}} \right)  \right] ,
\ea
where $a= 1.69$, $b=2.31$ and $c=0.46$  \cite{Dufaux:2006ee}.

Using the approximations in
{\bf Appendix. \ref{harmonicnumbers}} for the harmonic sums, we obtain
\begin{equation}
\label{X-sum3}
\sum_{j_{\ast}+1}^{j} 2 X_k^j \simeq \dfrac{8\alpha}{\pi}\sqrt{q_0} \left( \,   (j\pi )^{1/2}-( j_{\ast}\pi)^{1/2}  \,  \right) - \dfrac{12\alpha A_k^0}{\pi\sqrt{q_0}}\left( \,  (j\pi )^{1/6}-( j_{\ast}\pi)^{1/6} \, \right  ) \, .
\end{equation}
Finally,  by substituting $j \pi \rightarrow t$ after $j^{th}$ oscillation, the occupation number as a function of time is
\begin{equation}
\label{m2phi2n}
n_{j} \simeq \exp \left[ \dfrac{8\alpha}{\pi}\sqrt{q_0}\left( t^{1/2}- t_{\ast}^{1/2}\right) - \dfrac{12\alpha A_k^0}{\pi\sqrt{q_0}}\left( t^{1/6}- t_{\ast}^{1/6}\right)
 \right]  \prod_{j_{\ast}}^{j}( 2 \cos \Theta_\K ^j ) \, ,
\end{equation}
where  $\Theta_\K ^j$ is given by (\ref{phase-quadratic}) and $t_* =  j_*\pi $.

 \begin{figure}[!t]
\centerline{\includegraphics[scale=.3]{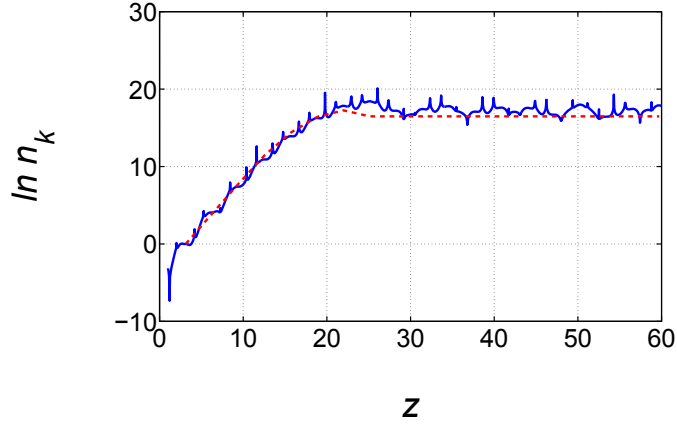}}
\caption{\label{ak20q13}
The logarithm of the occupation number as a function of  time for $A_k^0 = 20$ and $q_0=13$. The wavy (blue) curve shows the full numerical solution. The (red) dashed curve shows our analytical result (\ref{m2phi2n}). As expected from (\ref{t-end}), there is  no particle production after $t=23$.}
\end{figure}

Because of the expansion of the Universe, both $q$ and $A_\K$ decrease as $t$ increases, but
$A_\K$ decreases faster. On the other hand, the WKB approximation is valid for $2q -A > 2\sqrt{q}$ which quickly reduces to  $2q > 2 \sqrt{q}$. This condition is not satisfied for $q<1$. As a result, the expansion of the Universe spoils our approximation and that is  why the analytical solutions are not as precise as in the case of $\lambda \phi^4/4$ theory. As one can see from the numerical results in {\bf Fig. \ref{ak20q13}},   particle production  is stopped after some oscillations because of the expansion of the Universe.

In order to  interpret  this effect note that by reducing $A_k$ and $q$, the solutions of Mathieu equation converge to the stability bands. Therefore, in the expanding background when the $A_k-q$ curve crosses the stability bands,
the particle production switches off. From the stability/instability charts of Mathieu equation
\cite{abramowitz} one finds that preheating ends when
\begin{equation}
\label{t-end}
\dfrac{q_{0}}{t_{\mathrm{end}}}\simeq 0.8126 \left(1-\dfrac{A^0_k}{t^{4/3}_{\mathrm{end}}} \right).
\end{equation}
As a results, for initial conditions with $A_k^0 > 2q_0$, there can be particle production only if
$t_{\ast} > t_{\mathrm{end}}$. This does not take into account the effects of back-reaction
which will be studied in the next section.


As already mentioned,  expansion of the Universe dilutes the condensate of the produced particles  as well as
reducing $\phi(t)$ as the source of resonance. As a result the
resonance in an expanding Universe is expected  to be less efficient
than the non-expanding background.  More investigations shows, however,
that for some specific modes $\K$ the expansion of Universe can  actually enhance the preheating! There are two effects which can enhance particle production in an expanding background. First and less important is the $k^2/a^2$ term in preheating equation
(\ref{mathieu}). The expansion of the Universe with the effect of  $k^2 \rightarrow k^2/a(t)^2$
can reduce the energy cost of producing particle (but there is a trade-off between this effect and the  effects of diluting $\phi$-condensate and the reduction of the source term). Second and the more important is the effect of varying interference term in an expanding background.

We first focus on the former effect. The occupation number in a flat background is given by \cite{Dufaux:2006ee}
\ba
\label{Lev}
n_\K^j \simeq \exp \left[ 2j\left( 2\alpha\sqrt{q_0}-\dfrac{\alpha}{\sqrt{q_0}}A^0_k \right) \right] \left( 2 \cos \Theta_\K \right)^{2(j-1)} \, ,
\ea
where
\ba
\Theta_k = \sqrt{q}\left[a+b\dfrac{A_k}{2q}+c\left(1-\dfrac{A_k}{2q}\right) \ln \left(1-\dfrac{A_k}{2q}\right) \right] \, .
\ea

\begin{figure}
\includegraphics[scale=.25]{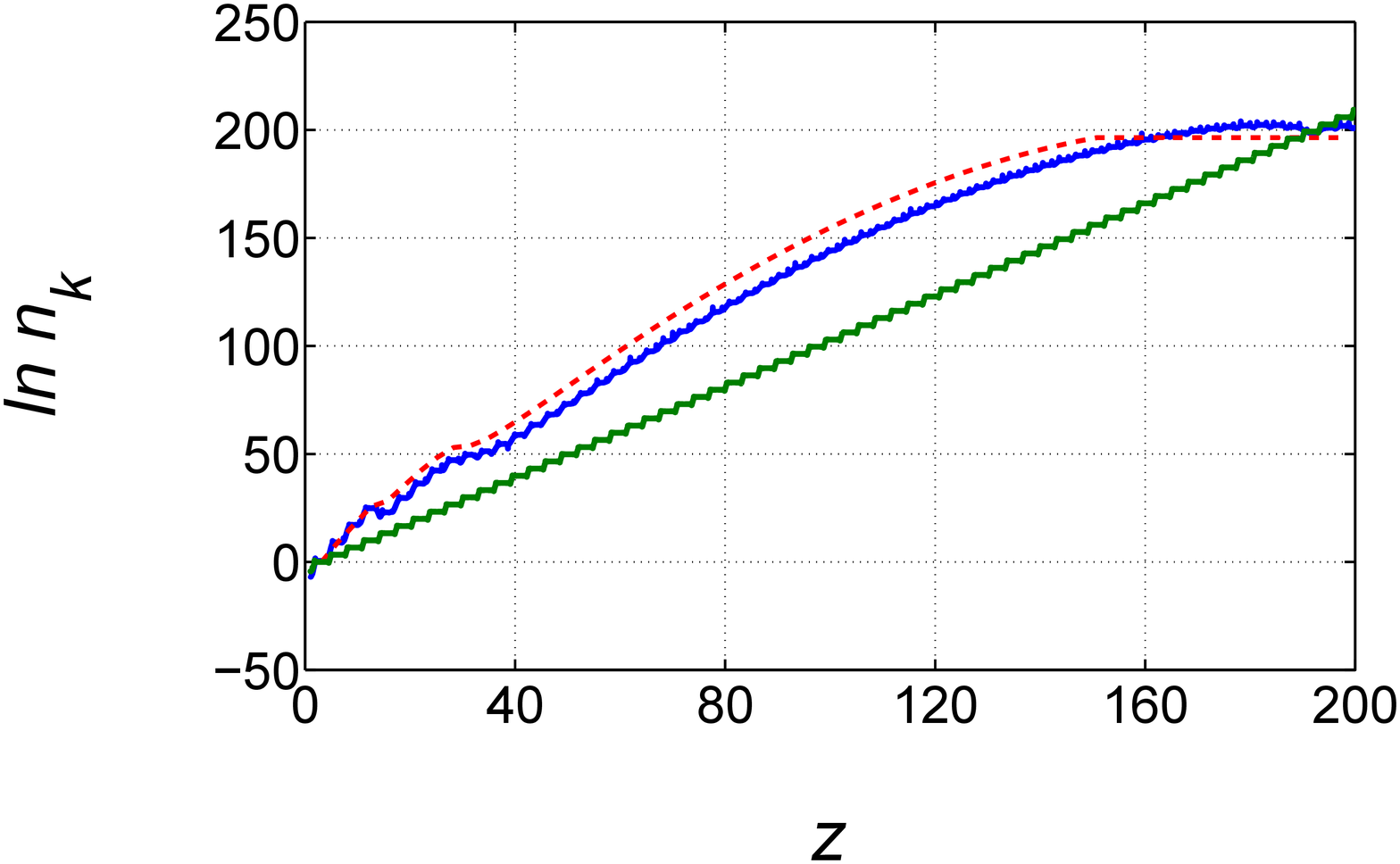}
\includegraphics[scale=.25]{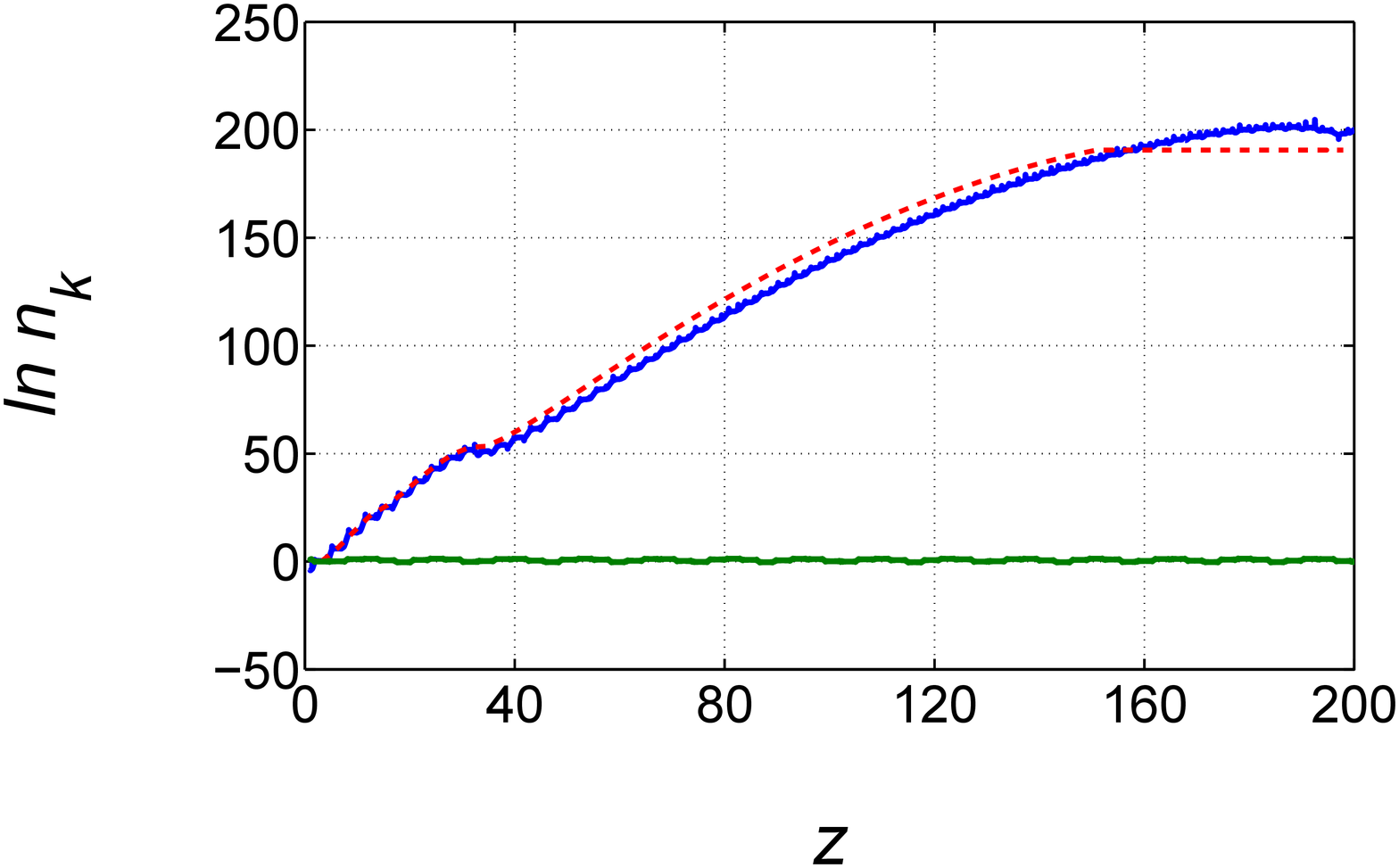}
\caption{\label{expnonexp}
Left: the logarithm of the occupation number as a function of time  for $A^0_k=230$ and $q_0 =125$. It shows that the expansion of the Universe can enhance the particle creation
by reducing the energy cost of particle production via $k^2 \rightarrow k^2/a^2(t)$. The  solid curve (blue) shows our full numerical result in an expanding Universe. The lower straight solid line  (green) shows numerical results for non-expanding universe. The dashed curve (red) shows our analytical results in an expanding background (\ref{m2phi2n}).
Right: logarithm of the occupation number  for $A^0_k=260$ and $q_0 =125$.
This figure shows that because of the time varying nature of the interference term, the solutions can escape from the stability bands. The  solid curve (blue) represents our full numerical result in an expanding Universe. The bottom  solid line (green) shows the numerical results for non-expanding Universe. The  dashed curve (red) shows our analytical solution in an expanding background (\ref{m2phi2n}).}
\end{figure}

As one can expect from the Floquet theorem,
the occupation number has a linear exponential growth  with time
(or number of oscillations) in a flat background. Compare this with our
result (\ref{m2phi2n}) where the occupation number grows with an exponent
which scales like $t^{1/2}$ or $j^{1/2}$. In general this leads to the
conclusion that at large $t$ the tachyonic resonance is less efficient
in the expanding background as compared to the flat background. For the intermediate
times the situation could be different.
To see this note that the  second term in the big bracket in (\ref{m2phi2n}),
which suppresses the particle creation, scales like $t^{1/6}$ in an expanding background
whereas it scales linearly with $t$ in a flat background. This difference in scaling results in an enhancement of particle creation  for some certain modes in an expanding background. However, after some oscillations the second terms in brackets in (\ref{m2phi2n}) and (\ref{Lev}) become negligible compared to the first terms in the corresponding brackets. Approximately this occurs for $t \sim 30 t_{\ast}$ and particle production in an expanding background becomes more and more inefficient compared to the static Universe. One can see this feature for $q_0 = 125$ and $A^0_k  = 230$ in {\bf Fig.\ref{expnonexp}},  showing that the particle production in an expanding background is more efficient for about first $60$  oscillations.

As mentioned above the interference term can play an  important role in relative enhancement of particle production in an expanding background. In the stability/instability chart of the Mathieu equation there are vast regions where there is no particle production due to destructive interference term.  However, as we described at the end of subsection \ref{non-zero-k}, in an expanding background the phase term is not static and is varying with time.
As such and as time advances, the solutions in the stability bands can escape from the stability bands.
Inside the tachyonic region, far from $A_k =2q$ line,  the effect of phase term is suppressed, but by going towards the  line $A_k=2q$ the stable regions are formed. So, in a
flat background there is no particle creation in this region but
as described above, in an expanding background, the solutions can escape
these stability bands, leading to an enhancement in particle creation (see the right graph in
{\bf Fig.\ref{expnonexp}}).

\section{Back-reaction Of $\chi$-particles and end of preheating}
\label{back-reaction}
So far we have neglected the back-reactions of the produced particles on classical background
$\phi$ and on $\chi$ particles occupation number. There are important effects which can change this simple picture.  We classify these effects by the form of their interactions.

First, trilinear interaction at one-loop Hartree approximation contributes as a source term in $\phi$ equation of motion. 
This leads to a non-zero
vacuum expectation value  for $\phi$ which is expected, because as one can see
from (\ref{V-eq}), the  minimum of the potential is located at a non-zero value of $\phi$.
Second, the self-interaction of $\chi$-particles can increase the effective mass of $\chi$-particles. This effect can make $\chi$-particles so heavy that they can not be produced through interaction with $\phi$, terminating the preheating.

\subsection{Back-reaction in $\lambda \phi^4$ theory}
Let us first ignore the effects of decay of inflaton field through trilinear interaction and consider the inflaton field as a background field. The preheating is complete if all the energy from the background field $\phi$ is
transferred into created $\chi$-particles. One can estimate the energy in the $\chi$ particles as%
\ba
\label{rho-chi-def}
\rho_{\chi} =\frac{1}{a^2(x)} \int\dfrac{\md ^3k }{(2\pi)^3} |\omega_k| n_k^{\chi}\ .
\ea%
The $1/a^2$ factor in front of the integral has appeared because we are working in the conformal frame and
that $\rho_{conf}=\rho_{comoving}/a^2$. Note also that $n_k^{\chi}=n^j_{\K}/a^2(x)$, with $n^j_{\K}$ given in
\eqref{nwithphase}. The integral over $k$ is cut off at $k^2_{max}\lesssim \lambda \tilde\varphi^2\ px $.
This upper bound comes from the fact that for a given time $x$ particle creation starts
for $x>x_{\ast}$, which in turn, recalling \eqref{j-star}, implies the bound mentioned above. %

In the conformal frame, at one loop Hartree approximation, the dispersion relation is%
\ba\label{disperssion}%
\omega^2_\K=\lambda \tilde\varphi^2\left(\kappa^2+px f(x)+3\frac{\lambda'}{\lambda}\frac{\langle \hat\chi^2\rangle}{\tilde\varphi^2} \right) ,
\ea%
and for our estimate of $\rho_\chi$ at this stage, we drop the last term
and justify this approximation later on in this subsection.
Moreover, for the period when the particle creation is more pronounced  one may also drop  the $\kappa^2$ term and approximate $\omega^2$ by $\lambda\tilde\varphi^2\ pxf(x)$ or
\ba\label{omegak-approx}%
\omega\simeq  \sqrt{\sigma a(x) |\varphi(x)|}\ .
\ea%
With this approximation
\ba\label{rho-chi-approx}
\rho_{\chi} =\frac{1}{a^4(x)} |\omega| \int\dfrac{\md ^3k }{(2\pi)^3} n_k^j\equiv \frac{1}{a^4(x)} n_{\chi} |\omega|
\ea%
where \footnote{Here $a=2.72$, as given in \eqref{a,b},  and  should not be mistaken with the scale factor $a(x)$.}
\ba\label{n-chi}%
n_\chi  \approx \int_0^{k_{max}} \frac{\md^3k}{(2\pi)^3}\ \exp \left( \dfrac{4a \sqrt{p}}{3T} \, \left(x^{3/2} - x_{\ast}^{3/2} \right) - \dfrac{4b'\kappa^2}{\sqrt{p}T} \left( x^{1/2} -x_{\ast}^{1/2} \right) \right)\
\ea
is the total number of $\chi$ particles produced. In order to perform the above integral we note that it could be written in the form
\ba\label{n-chi-k3k2}
n_\chi \approx \dfrac{e^{\zeta \sqrt{p}x^{3/2}}}{2\pi^2 }(\lambda \tilde{\varphi}^2)^{3/2}\int_{0}^{\sqrt{px}} \md \kappa \kappa^2 e^{A \kappa^3 - B \kappa^2 }\ ,
\ea
where $\zeta = 4 a /(3 T)\simeq 0.5$, $A=(12b'-4a)/3pT$ and
$B=4b'\sqrt{x}/\sqrt{p}T$. This integral cannot be calculated analytically
and one should approximate it. Comparing the two terms in exponent of the integrand
\ba
\dfrac{A\kappa^3}{B \kappa^2}= \dfrac{(3b' -a)}{3b'} \dfrac{\kappa}{\sqrt{px}}
\simeq 0.7 \dfrac{\kappa}{\sqrt{px}}\ ,
\ea
we see that for the range of integration $\kappa < \sqrt{px}$ one can
neglect term $A\kappa^3$ and just take $-Bk^2$ term, reducing the integral
to an incomplete Gaussian integral. For large values of $x$, which we are interested in, i.e. for $\sqrt{p}x^{3/2}\gtrsim T/b'\approx 2.6$, the integral can be computed leading to
\ba\label{n-chi-final}%
n_\chi \approx \frac{1}{64}\left(\frac{\sqrt{p} T}{\pi b'}\right)^{3/2}\ (\lambda \tilde{\varphi}^2)^{3/2} x^{-3/4} {e^{\zeta \sqrt{p}x^{3/2}}}\ .
\ea

Preheating completes at $x_{\mathrm{cop}}$, where%
\ba
\rho_{\chi}(x_{\mathrm{cop}}) a^3(x_{\mathrm{cop}}) \sim \rho_{\phi}^0=\dfrac{\lambda}{4}\tilde{\varphi}^4 \, .
\ea
which happens when
\ba\label{x-cop}%
x_{\mathrm{cop}}^{-5/4}e^{\zeta
\sqrt{p}x_{\mathrm{cop}}^{3/2}}\approx \frac{16}{\lambda p^{5/4}}
\left(\frac{\pi b'}{T}\right)^{3/2}\
\sqrt{\frac{2\pi}{3}}\frac{\tilde\varphi}{M_P}\approx
\frac{3.1}{\lambda p^{5/4}}\ . \ea As one expects by increasing $p$
preheating shuts off sooner. This is reasonable, since the bigger
the value of $p$, the stronger is the trilinear interaction which
results in a more efficient $\chi$ particle production and a
stronger back-reaction effect. As an example, with $p=0.05$ and
$\lambda=10^{-14}$ one finds $x_{\mathrm{cop}}\simeq 52.5$ which is
about $7$ oscillations.

As the density of $\chi$ particle grows and is seen from
\eqref{disperssion}, the effective mass of $\chi$ particles also
grows and production of them becomes more costly and eventually
terminate the preheating. This happens at $x_{1}$, and at first loop
Hartree approximation it is
when%
\ba\label{eop1}%
3{\lambda'}{\langle \hat\chi^2\rangle}\big|_{x_{1}}\approx \lambda
p\
x_{1} {\tilde\varphi^2}. %
\ea %
On the other hand, the produced $\chi$-particles also back-react on
the dynamics of the inflation field $\phi(t)$ and may cause
preheating to stop before completion, making our preheating model
inefficient. To check when this can happen we note that back
reaction of $\chi$-particles at one loop Hartree approximation
level, modifies the $\varphi$ equation \eqref{phi-eq} to
\ba%
\varphi''+\lambda\varphi^3+\frac{\sigma a(x)}{2} {\langle
\hat\chi^2\rangle}=0\ .
\ea%
The back reaction of $\chi$-particles on $\varphi$ becomes important
at $x_{2}$, when
\ba\label{eop2}%
\sigma \left(a(x){\langle \hat\chi^2}\rangle\right)\big|_{x_{2}}
\approx 2\lambda\tilde\varphi^3\ .
\ea%
Whichever of the two stoping mechanisms happens first marks the end
of preheating. We will denote this time by $x_{\mathrm{eop}}$, where
$x_{\mathrm{eop}}=min(x_{1}, x_{2})$. The condition of having a
successful preheating is then $x_{\mathrm{eop}}\gtrsim
x_{\mathrm{cop}}$.

To evaluate the $x_{1}$ and $x_{2}$ we note
that within our approximations%
\ba\label{chi-VEV}%
{\langle \hat\chi^2\rangle}= \int\dfrac{\md ^3k }{(2\pi)^3} \frac{n_k^{j}}{|\omega_k|}\approx \frac{n_{\chi}}{|\omega|}\ ,
\ea%
with $|\omega|$ given in \eqref{omegak-approx}. This leads to%
\begin{subequations}\label{eop1,2-eqs}%
\begin{align}%
 n_\chi(x_{1}) &\approx \frac{1}{3\lambda'}\ \left(p\
\lambda \
x_{1}\right)^{3/2}\tilde\varphi^3\ ,\\
  n_\chi(x_{2})
&\approx 2\lambda\ \left(p\ \lambda  \
x_{2}\right)^{-1/2}\tilde\varphi^3\ .\
\end{align}
\end{subequations}%
Using $n_\chi$  given in \eqref{n-chi-final} we end up with%
\begin{subequations}\label{eop1,2-eqs-simplified}%
\begin{align}%
x_{1}^{-9/4}e^{\zeta \sqrt{p}x_{1}^{3/2}}&\approx
\frac{64}{3\lambda' } \left(\frac{\pi b'}{T}\right)^{3/2}\ p^{3/4}
\simeq
\frac{28.51}{\lambda'} p^{3/4}, \\
x_{2}^{-1/4}e^{\zeta \sqrt{p}x_{2}^{3/2}} &\approx
\frac{128}{\lambda} \left(\frac{\pi b'}{T}\right)^{3/2}\ p^{-5/4}
\simeq\frac{171}{\lambda p^{5/4}}.
\end{align}
\end{subequations}%

As we see  $x_1$ depends on $\lambda'$ whereas $x_2$ depends on
$\lambda$. For $\lambda\sim \lambda'\sim 10^{-14}$ and with
$p=0.05$, we find that $x_1\simeq x_2\simeq 52.5$. This means that
for this choice of parameters $x_{\mathrm{eop}}\simeq 52.5$ which is
very close to the completion of preheating time $x_{\mathrm{cop}}$
and hence we expect an efficient preheating.

Given the above expressions one may ask  for which range of
parameters $p,\ \lambda,\ \lambda'$  the ``efficient preheating''
condition $x_{\mathrm{eop}}\gtrsim x_{\mathrm{cop}}$ is satisfied. The
comparison between $x_2$ and $x_{\mathrm{cop}}$ is somewhat
straightforward, noting that in the range of parameters that we are
mainly interested in, $n_\chi/\omega$ is a monotonic function of $x$
and hence if $a(x_2)<8$ (or equivalently $x_2\lesssim 55$)
$x_2<x_{\mathrm{cop}}$ and vice-versa. Moreover, one can argue that
either $x_1^2>U>x_2^2$ with $x_1x_2>U$, or $x_1^2<U<x_2^2$ with
$x_1x_2<U$, where $U\equiv 6\lambda'/(\lambda p^2)$. In the former case,
the $x_{\mathrm{eop}}\gtrsim x_{\mathrm{cop}}$ condition roughly boils down
to $500p^2\frac{\lambda}{\lambda'}\gtrsim 1$. In the latter case,
when $x_{\mathrm{eop}}=x_1$, one may again show that a similar bound
holds. This condition together with \eqref{cond1} specifies the
range of parameters for which we have a simple slow-roll
$\lambda\phi^4$ inflation as well as efficient preheating:
\ba\label{lambda-range}%
2\times 10^{-3}\lesssim p^2\frac{\lambda}{\lambda'}\lesssim 1\ .
\ea%
For typical  $\lambda\sim \lambda'$ values, this leads to $2\times
10^{-3}\lesssim p^2\lesssim 1$.

It is also notable that  ${\sigma}{a(x)} {\langle
\hat\chi^2\rangle}\approx \frac{\sqrt{\sigma a(x)}\
n_\chi}{\sqrt{|\varphi|}}$ term shifts the minimum of the potential
of $\varphi$ from $\varphi=0$ and we expect the end point of
preheating to be not far from the $\varphi_{min}$. As a rough
estimate of this minimum value (assuming that $x_1\thicksim x_2$),
at this approximation level, is $\varphi_{min}^2\approx
a^2(x_2)\sigma^2/(6\lambda\lambda')$. For our estimates, where
$a(x_2)\sim 8$, this rough estimate is not far from the the global
minimum of the potential \eqref{V-eq} which is at
$\phi_0^2=\sigma^2/(2\lambda\lambda')$.

Before ending this subsection let us briefly discuss the production of $\varphi$ particles and re-scattering of $\chi$-particles in the theory with potential \eqref{V-eq}. At tree level the $\varphi_\K$ modes will be sourced by a term like $\lambda \langle \varphi^2\rangle$. This term being positive and periodic, following the discussions of
\cite{Greene:1997fu}, leads to stochastic resonant production of $\varphi_\K$ modes, which at large $x$, has an average exponential growth in $x$. This effect compared to the tachyonic resonant production of $\chi$ particles is very small and one may safely conclude that during preheating mainly $\chi$-particles are produced.

At one loop level there is a contribution to the equation of motion for the $\hat\chi_\K$ mode proportional to
$\lambda \sigma^2\varphi^2/k^2\propto \lambda^2 p^2 (\frac{M_P}{k})^2\ f^2(x)$. This, unlike the tachyonic source term, $px f(x)$, is always positive. Nonetheless being proportional to $\lambda^2 p^2$, this one loop contribution, is too small compared to the tree level $px f(x)$ term.
With the trilinear $\phi\chi^2$ coupling, produced $\chi$ particles will not back-react on the the production of $\varphi_\K$ modes, at first loop level beyond the Hartree approximation, which we have already discussed. This is in contrast with the four-leg $\phi^2\chi^2$ interaction, where a non-zero $\langle \chi^2\rangle$ contribute to the equation of motion of $\varphi_\K$.

\subsection{Back-reaction in $m^2 \phi^2$-theory with trilinear interaction}

The important point in this case is that following  (\ref{t-end}) and discussions leading to it,
we note that in the $m^2 \phi^2$ theory  there is a definite time, $t_\mathrm{end}$, after which there is no particle production at all. Therefore, to have an efficient preheating the energy transfer form
the $\phi$ background into $\chi$ particles should happen before this time.

As in $\lambda \phi^4/4$ case lets us estimate the time at which the energy transferred  into the $\chi$ particles becomes comparable to the background energy.
The energy density of $\chi$-particles after time $t$  is
\ba\label{rhox}
\rho_{\chi} = \int\dfrac{\md ^3k }{(2\pi)^3} |\omega_k| n_k^{\chi}=\frac{1}{a^3(t)}\int\dfrac{\md ^3k }{(2\pi)^3}
|\omega_k| n_j\ ,
%
\ea
where $n_j$ is given in (\ref{m2phi2n}) and from potential \eqref{m2phi2-potential} one reads that at one loop Hartree approximation
\ba\label{omega2-m2phi2}%
\omega^2_k=\frac{k^2}{a^2(t)}+\frac{\sigma \varphi(t)}{(a(t))^{3/2}}+\frac{3\lambda}{a^3(t)}\langle \hat\chi^2\rangle =k^2t^{-4/3}+\sigma t^{-1}\varphi(t)+3\lambda t^{-2}\langle \hat\chi^2\rangle\ .
\ea%
with $\varphi(t)=\Phi_0\sin mt$. The main contribution to the integral \eqref{rhox} comes from the period when the tachyonic resonance is at work.
In this period one may drop $k^2/a^2$ term in \eqref{omega2-m2phi2}. For the current estimation we also drop the back-reaction of $\chi$-particles, to which we will return later. Therefore,
\ba\label{omega-approx-m2phi2}%
\omega^2\simeq \frac{\sigma\varphi (t)}{a^{3/2}}=\sigma t^{-1} \varphi(t)\ .
\ea%
Inserting \eqref{omega-approx-m2phi2} into \eqref{rhox} we find
\ba\label{rhox-approx-m2phi2}%
\rho_\chi\approx\frac{1}{a^3} |\omega| \int \frac{\md^3k}{(2\pi)^3}\
n_j\equiv \frac{1}{a^3} |\omega| n_\chi\ ,
\ea%
where
\ba\label{n-chi-m2phi2}%
n_\chi\approx \int_0^{k_{max}}  \frac{\md^3k}{(2\pi)^3}\ \exp \left[\dfrac{8\alpha}{\pi}\sqrt{q_0}\left( t^{1/2}- t_{\ast}^{1/2}\right) - \dfrac{12\alpha A_k^0}{\pi\sqrt{q_0}}\left( t^{1/6}- t_{\ast}^{1/6}\right)\right]
\ea%
In the above, as in the $\lambda\phi^4$ case, the upper bound on
$k$, $k_{max}$, comes from the fact that particle creation becomes
efficient for $t>t_{\ast}=\pi j_{\ast}$ with $j_\ast$ given
in \eqref{j-star-m2phi2}, that is  $k_{max}\lesssim m\sqrt{\frac{q_0}{2}} t^{1/6}$. To perform the integral we note that it has the same form as in \eqref{n-chi-k3k2}, with a cubic and a quadratic term in the exponent, where now
$A=\frac{32\sqrt{2}\alpha}{\pi m^3 q_0},\ B=\frac{48\alpha}{\pi m^2\sqrt{q_0}} t^{1/6}$ and hence $Ak/B\lesssim 2/3$.
One may again drop the $Ak^3$ term and approximate the integral by an incomplete Gaussian integral:
\ba\label{n-chi-approximated}%
n_\chi\approx   \dfrac{e^{\zeta \sqrt{q_0 t}}} {2\pi^2 }  \int_{0}^{k_{max}} \md k\ k^2 e^{-Bk^2}\simeq
\frac{m^3 q_0^{3/4}}{512(3\alpha)^{3/2}} t^{-1/4}e^{\zeta \sqrt{q_0 t}}
\ea
where $\zeta = 8 \alpha / \pi \simeq 2$ and in the last step we have approximated the integral for $ q_0 t \gtrsim 1$.

Preheating is complete at  $t_{cop}$ when the energy of $\chi$-particles becomes comparable to the energy in the the inflaton condensate at the beginning of preheating in the same comoving volume, i.e.
\ba\label{t-cop}
{\rho_{\chi} a^3(t)}\Big|_{t_{cop}}= \left(n_\chi |\omega|\right)\Big|_{t_{cop}}\approx \frac12 m^2\Phi_0^2\ .
\ea
Using the expressions given above we obtain
\ba\label{t-cop-II}
t_{cop}^{-3/4}e^{\zeta \sqrt{q_0 t_{cop}}} \approx 256\sqrt{2} (3\alpha)^{3/2} q_0^{-5/4}
\left( \dfrac{~\Phi_0}{m} \right)^2\simeq 1474\ q_0^{-5/4}
\left(\dfrac{~\Phi_0}{m} \right)^2 .
\ea
For $q_0=50$ and $\Phi_0/m\sim 10^5$, the above is satisfied for $t_{cop}\simeq 3.5$ which is slightly more than one period time (which is $t=\pi$). In order to have an efficient preheating, we should demand that the preheating completes before the particle creation stops. The latter happens at $t_\mathrm{end}$ given in (\ref{t-end}). $t_{\mathrm{end}}> t_{cop}$ imposes a lower bound
on $q_0$. For $\Phi_0 / m \sim 10^5$, we find
\ba
q_0 > q_0^{c} \simeq 13 \, .
\ea
Noting  that $q_0= 2\sigma \Phi_0 /m^2$, this can be used to impose a lower bound
on $\sigma$, the scale involved in trilinear coupling, $\sigma/m > 6.5\ m/\Phi_0\simeq 6.5 \times 10^{-5}$.

As $\langle \chi^2\rangle$ increases $\chi$-particle effective mass increases, \textit{cf.}\eqref{omega2-m2phi2}, which in turn can stop preheating. One needs to also verify that this happens after the completion of preheating, that is
\ba\label{good-preheat-m2phi2}
3\lambda t^{-2} \langle\hat\chi^2 \rangle  \lesssim{\sigma t^{-1}\varphi(t)}
\ea
at $t_{cop}$. Noting that
\ba\label{chi2-m2-phi2}%
\langle\hat\chi^2 \rangle \approx n_\chi\frac{1}{|\omega|}\ ,
\ea
with $n_\chi$ given in \eqref{n-chi-approximated}, $\lambda \langle\hat\chi^2 \rangle\approx \sigma t \varphi(t)$ happens when $(q_0 t)^{-3/4} e^{\zeta \sqrt{q_0 t}}\simeq \frac{512}{3\lambda} (\frac{3\alpha}{2})^{3/2}$. The condition \eqref{good-preheat-m2phi2} is satisfied if $\lambda\lesssim 0.17 (\frac{m}{\Phi_0})^2\ q_0^2=0.67 \left(\frac{\sigma}{m}\right)^2$. Noting the condition for positivity of the potential \eqref{m2phi2-potential}, an
efficient preheating scenario  happens if $\lambda$ is in the very tight range:
\ba\label{lambda-range-m2phi2}%
0.5 \left(\frac{\sigma}{m}\right)^2\leq\lambda \lesssim 0.67 \left(\frac{\sigma}{m}\right)^2\ .
\ea
For $\sigma/m\sim 10^{-4}$  (when $q_0=20$) that is, $ 5\times 10^{-9}\leq \lambda\lesssim 6.7\times 10^{-9}$.

In the $m^2\phi^2$ case, with the potential \eqref{m2phi2-potential}, at one loop Hartree approximation level the
equation of motion for the background $\phi$ field is modified as%
\ba\label{modified-phi-eq}%
\ddot\varphi +m^2\varphi+\frac{1}{2a^{3/2}}\sigma \langle\hat\chi^2 \rangle=0\ ,
\ea
where $a^{3/2}=t,\ \varphi=a^{3/2}\Phi$ and $\hat\chi=a^{3/2}\chi$. Using \eqref{chi2-m2-phi2} and \eqref{omega-approx-m2phi2}, we observe that the force term in \eqref{modified-phi-eq} {at the completion of preheating time} can vanish for a non-zero $\varphi$, $\varphi_{min}$:
$m^2\varphi_{min}\sim \sigma t^{-1} n_\chi/2 \omega\big|_{t=t_{cop}}$ which upon using \eqref{t-cop} we obtain%
\ba
m^2\varphi_{min}\sim \sigma t^{-1}\frac{\frac12 m^2\Phi_0^2}{2\sigma t^{-1} \varphi_{min}}=\frac{m^2\Phi_0^2}{4 \varphi_{min}}
\ea
or $\varphi_{min}\sim \frac12\Phi_0$. It is interesting that our rough estimates and considering the one loop Hartree approximation reproduces the order of magnitudes of the minimum obtained in numerical analysis of \cite{Dufaux:2006ee} (see FIG3 in Ref.\cite{Dufaux:2006ee}).


\section{Summary}
\label{summary}
In this paper we have studied tachyonic resonance preheating via trilinear
interaction in an expanding background. Because of the  three-leg
interaction the conformal symmetry in $\lambda \phi^4/4$ inflationary theory is broken. This induces a non-periodic
source term in resonant $\chi$ field  equation.
Interestingly, one observes that the particle creation has a non-linear exponential enhancement with the leading exponent $\sim x^{3/2}$. This is in contrast to particle creation via parametric resonance from the four-legs interaction in this theory which preserves the conformal invariance and a linear exponential growth of particle creation
is obtained.
 Besides the non-linear exponential growth, the interference term obtained from the accumulation of the phase term in non-tachyonic scattering regions  has a very non-trivial behavior in an expanding background. It is shown that there are ``semi stability bands'' where
$\partial_j \Theta_\K^{j_s} = \cos \Theta_\K^{j_s}=0$ for some oscillations $j_s$. As a result, the particle creation for the corresponding modes are highly suppressed. However, due to time varying nature of the phase term, the semi-stability bands are washed out as time goes by.

The tachyonic resonance preheating in an expanding background for $m^2 \phi^2/2$ theory
was also studied. In general the expansion of the Universe suppresses the particle creation.
However, due to time varying nature of the interference term, the expansion of the Universe can actually enhance the particle creation for certain modes.

We studied in some detail, the back reaction of preheat $\chi$-particles on the dynamics of preheating and the background inflaton field $\phi$. As we argued demanding an efficient preheating (i.e. demanding that particle production ends not before most of the energy in the background $\varphi$ field is transferred into the $\chi$-particles) imposes strong bounds on the parameters of the potential in both $\lambda\phi^4$ and $m^2\phi^2$ cases.

For typical values of the parameters of the potential for $\lambda\phi^4$ theory preheating is complete and lasts for $\lesssim 10$ oscillations while for $m^2\phi^2$ in few oscillations the energy transfer to $\chi$-particles seems to be complete.


 \section*{Acknowledgments}
We are deeply grateful to late Lev Kofman for stimulating discussions and for bringing Ref. \cite{Dufaux:2006ee} into our attention which started this project. To our grate sadness that he passed away
while this work was almost finished.
We also would like to thank Bruce Bassett and Robert Brandenberger for useful discussions and  Hesam Moosavimehr
for computational assistances. H. F. would like to thank CITA and
Perimeter Institute for hospitality where this work started. A. A. would like to thank  IPM and ``Bonyad Nokhbegan Iran'' for partial support.


\appendix

\section{Properties of Jacobi elliptic cosine functions $cn(x|m)$}
\label{acn}

Here we briefly review the properties of the Jacobi elliptic cosine function $cn (z|m)$. Note that we follow the convention of \cite{abramowitz} which differs from the convention of \cite{Greene:1997fu}.
 A useful representation of the Jacobi elliptic cosine functions is
\cite{wolfram}
\begin{equation}
\label{jacobicn}
cn(z | m) = \dfrac{2 \pi }{\sqrt{m} K(m)} \sum_{n=0}^{\infty} \dfrac{q(m)^{n+1/2}}{1+q(m)^{2n+1}} \cos \, \left( (2n+1)\dfrac{\pi z}{2 K(m)} \right)  \, ,
\end{equation}
where
$$q(m) = \exp \left( -\dfrac{\pi K(1-m)}{K(m)} \right)$$
and $K(m)$ is the complete elliptic integral of the first kind.

The Jacobi elliptic cosine is periodic with period $T=4K(m)$ and has zeros at
\begin{equation}
\label{turningpoint}
x_{j}^{-} = (j - \dfrac{3}{4}) T  , \qquad x_{j}^{+} = (j + \dfrac{1}{4}) T.
\end{equation}
The minima and maxima are at
\begin{equation}
x_{j}^{max} = ( j - 1) T,
\qquad
x_{j}^{min} = (j - \dfrac{1}{2})T .
\end{equation}
\\ For the special case of $ m = 1/2$ this formula reduces to
\begin{equation}
f(x) = cn(x | \frac{1}{2}) = \dfrac{2 \sqrt{2}\pi }{ K(\frac{1}{2})} \sum_{n=0}^{\infty} \dfrac{\mathrm{e}^{-\pi(n+1/2)}}{1+\mathrm{e}^{-\pi(2n+1)}} \cos \, \left( (2n+1)\dfrac{\pi z}{2 K(\frac{1}{2})} \right)
\end{equation}
As can be seen from above, $f(x)$ is even and periodic with the periodicity
 $T = 4 K(\frac{1}{2}) = \frac{\Gamma^2(1/4)}{\sqrt{\pi}} \approx 7.416$.


\section{Approximations for generalized harmonic numbers }
\label{harmonicnumbers}
Here we present some useful formula for the sums of the harmonic numbers which are used
to calculate the sum over $j$ for the occupation number in (\ref{X-sum}),
(\ref{X-sum2})  and (\ref{X-sum3}).

With straightforward algebra one can check that
\begin{equation}
\sum _{i = 1 }^{j} \sqrt{i} \simeq \dfrac{2}{3} j^{3/2} + \dfrac{1}{2} j^{1/2} - \dfrac{1}{6}
\end{equation}
\begin{equation}
\label{1/j}
\sum _{i = 1 }^{j} \dfrac{1}{\sqrt{i}} \simeq 2 j^{1/2} + \dfrac{1}{2} j^{-1/2} - \dfrac{3}{2}
\end{equation}
and
\begin{equation}
\label{2j-1}
\sum _{i = 1 }^{j} \sqrt{2i - 1} \simeq \dfrac{2\sqrt{2}}{3} j^{3/2}
\end{equation}
which are numerically confirmed to have a very good accuracy.

One can generalized these approximations  for arbitrary power  $ s\neq -1$
\begin{equation}
\sum _{i = 1 }^{j} i^s \simeq \dfrac{1}{s+1} j^{s+1} +\dfrac{1}{2} j^s + \dfrac{s-1}{2s+2} \, .
\end{equation}


\section{Approximation for calculating  $\int ^ {x^{max}_{j} +T/4} _{x^{max}_{j}-T/4} \sqrt{xf(x)} \, \mathrm{d}x$}
\label{approximation}
In this section we would like  to examine the approximation used in calculating the integral
in (\ref{X-int}):
\begin{equation}
\int ^ {x^{max}_{j} +T/4} _{x^{max}_{j}-T/4} \sqrt{xf(x)} \, \mathrm{d}x \simeq \sqrt{x^{max}_{j}}\int _{-\frac{T}{4}}^{+\frac{T}{4}} \sqrt{f(x)} \, \mathrm{d}x \, .
\end{equation}
Defining $x = x_{j}^{max} + u$, one finds

\begin{equation*}
\int ^ {x^{max}_{j} +T/4} _{x^{max}_{j}-T/4} \sqrt{xf(x)} \, \mathrm{d}x = \sqrt{x^{max}_{j}}\int _{-T/4}^{T/4} \sqrt{1+\dfrac{u}{x_{j}^{max}}}\sqrt{f(x_{j}^{max}+u)} \, \mathrm{d}u
\end{equation*}
\begin{equation}
\qquad \qquad \qquad= \sqrt{x^{max}_{j}}\int _{-T/4}^{T/4} \sqrt{1+\dfrac{u}{x_{j}^{max}}}\sqrt{f(u)} \, \mathrm{d}u \ .
\end{equation}
In the last step we have used the relation $f(x_{j}^{max}+u) = f(x)$ (see Appendix \ref{acn}). Expanding the first square root in the integral and noting that $f(x)$ is an even function one finds
\begin{equation}
\int ^ {x^{max}_{j} +T/4} _{x^{max}_{j}-T/4} \sqrt{xf(x)} \, \mathrm{d}x = \sqrt{x^{max}_{j}}\int _{-T/4}^{T/4} \sqrt{f(u)} \, \mathrm{d}u \,\,\,\left(1+\mathcal{O}\left( \dfrac{1}{16 ( j - 1)^2}\right)  \right).
\end{equation}
For our case, we need $j > 1$, so the first correction can lead to maximum $6\% $  error.  Since the higher corrections are inversely related to square of $j-1$, the errors decay very quickly.  For example for $j=4$ the error is less than $1 \%$.

\section{Approximation for calculating $\int_{x_{j-1}^{+}(r)}^{x_{j}^{-}(r)} \sqrt{r + f(x)} \mathrm{d} x $ for $ 0 <  r <1 $}
\label{r+jacobicn}
Here we demonstrate the approximations used to calculate the integral in (\ref{X-int2})
and (\ref{Theta-int2})

From \ref{jacobicn} one finds that
\begin{equation}
cn(x + x _{j}^{max}|m) = cn(x|m) \qquad \mathrm{and} \qquad cn(x + x _{j}^{min}|m) = - cn(x|m)
\end{equation}
 The function $ r + f(x)$ with $0 < r <1$ has zeros at
\begin{equation*}
x_{j}^{-}(r) = x_{j}^{max}(r) + K \left(\arccos(-r) | \frac{1}{2} \right) = x_{j}^{min}(r) - 2K \left(\frac{1}{2} \right) + K \left(\arccos(-r)|\frac{1}{2} \right)
\end{equation*}
\begin{equation}
x_{j}^{+}(r) = x_{j +1}^{max}(r) - K \left(\arccos(-r)|\frac{1}{2} \right) = x_{j}^{min}(r) + 2K \left(\frac{1}{2} \right) - K \left(\arccos(-r)|\frac{1}{2} \right)
\end{equation}
where $K(x|m)$ is the incomplete elliptic integral of the first kind with parameter $m$ and amplitude $x$ \cite{abramowitz}.
 We propose following approximations:
 \begin{equation}
 \label{adiabaticint}
\int_{x_{j-1}^{+}(r)}^{x_{j}^{-}(r)} \sqrt{r + f(x)} \mathrm{d} x \simeq  a + b\, r  \, ,
\end{equation}
with $a = \sqrt{\frac{\pi}{2}} \dfrac{\Gamma \left( \frac{3}{8} \right)}{\Gamma \left( \frac{7}{8} \right)}=2.72\ $, $b = 3.748$ and
\begin{equation}
\label{tachyonicint}
\int_{x_{j}^{-}(r)}^{x_{j}^{+}(r)} \sqrt{ \vert r + f(x) \vert } \mathrm{d} x \simeq a - b'\, r  \, ,
\end{equation}
with $b'= 2.864$. The fact that $b'>0$ has the important consequence  that the occupation number $n_\K\neq0$ is more suppressed for higher values of $k$, see subsection \ref{non-zero-k} for the details.

Both of these approximations are verified numerically to a good accuracy. Of course, to have
higher accuracies one can keep higher powers of $r$ in above expansions.



\end{document}